\pdfoutput=1

\documentclass[fontsize=9pt, DIV=calc, a4paper, twocolumn, headings=standardclasses]{scrartcl}

\usepackage[utf8]{inputenc}
\usepackage{lmodern}
\usepackage[english]{babel}
\usepackage[activate={true, nocompatibility}, final, tracking=true, kerning=true, factor=1100, stretch=10, shrink=10]{microtype}
\usepackage[group-separator={,}]{siunitx}
\usepackage{booktabs, graphicx, amsmath, amssymb, physics, enumitem}
\usepackage[format=plain, labelfont=bf]{caption}
\usepackage[auth-lg, affil-it]{authblk}
\usepackage[compress, numbers, sort]{natbib}
\usepackage[hidelinks]{hyperref}

\setlist{itemsep=2pt,topsep=2pt,parsep=0pt,partopsep=0pt,leftmargin=*}

\makeatletter
\doforeachtocfile{\unsettoc{\@currext}{onecolumn}}
\makeatother

\newcommand{\program}{PolyHoop}
\title{\program{}: Soft particle and tissue dynamics with topological transitions}

\author[$1,2,*$]{Roman Vetter}
\author[$1,2$]{Steve V. M. Runser}
\author[$1,2$]{Dagmar Iber}

\affil[$1$]{Department of Biosystems Science and Engineering, ETH Z\"{u}rich, Mattenstrasse 26, 4058 Basel, Switzerland}
\affil[$2$]{Swiss Institute of Bioinformatics, Mattenstrasse 26, 4058 Basel, Switzerland}

\date{26 July 2023}

\begin{document}

\maketitle

\noindent\textbf{We present \program{}, a lightweight standalone C++ implementation of a mechanical model to simulate the dynamics of soft particles and cellular tissues in two dimensions. With only few geometrical and physical parameters, \program{} is capable of simulating a wide range of particulate soft matter systems: from biological cells and tissues to vesicles, bubbles, foams, emulsions, and other amorphous materials. The soft particles or cells are represented by continuously remodeling, non-convex, high-resolution polygons that can undergo growth, division, fusion, aggregation, and separation. With \program{}, a tissue or foam consisting of a million cells with high spatial resolution can be simulated on conventional laptop computers.}

\vskip\baselineskip
\noindent\textbf{Keywords:} soft particle, foam, bubble, cell, tissue, polygon

\noindent\textbf{$^*$Correspondence:} vetterro@ethz.ch

\tableofcontents

\section{Introduction}

A number of two-dimensional mechanical systems occurring in Nature and our daily life can be described by the dynamics of elastic, tensile hoops marking the boundaries of domains consisting of fluidic materials. The most prominent example, perhaps, is the packing of biological cells in monolayer tissues, in which cell membranes surrounding the cytoplasm can effectively be modeled as constrained adhesive polygons under tension \cite{Kawasaki:1989, Weliky:1990, Graner:1993, Nagai:2001, Farhadifar:2007, Hufnagel:2007, Vetter:2019}. Major progress in the physical understanding of epithelial dynamics has been made in recent years with 2D computer simulations \cite{Bi:2015, Kim:2021}. Foams and froths, which consist of gas bubbles enclosed by a liquid phase, are another prime example. Depending on the degree of wetness, the gas chambers in foams can exhibit a wealth of shapes and arrangements, which has made them a field of intense study at the interface of geometry and physics over decades \cite{Weaire:1992, Weaire:2007, Weaire:2009}, also computationally.

Computer simulations of such systems started mainly in the 1980s with vertex models \cite{Odell:1981, Kawasaki:1989, Weliky:1990}, followed by a program named \textit{2D-FROTH} representing dry froths and foams with curved bubble boundaries \cite{Kermode:1990}---a feature also particularly important in tissue biology \cite{Ishimoto:2014, Perrone:2016, Boromand:2019}. These early implementations used shared polygonal boundaries between neighboring cells. More topological freedom in the form of individual, free boundaries was introduced with a C program named \textit{PLAT} shortly after \cite{Bolton:1992,PLAT:1996}. In the 2000s and 2010s, several new two-dimensional mechanical models were developed to simulate a variety of phenomena including tissue growth and morphogenesis, cell migration and aggregation, soft particle packing, and many more. We review the history of developed computer programs that continued along the path of polygonal representations of fluidic domains in Table~\ref{tab:1}. About half of these computational models were made open-source, published under various licenses, with a tendency in the last decade toward more openness, but also more copyleft.

\begin{table*}
\centering
\small
\caption{\textbf{Overview of related computer programs.} Only models in two spatial dimensions and with geometrically represented, curved cell/bubble boundaries are included. Line numbers, runtimes and cell counts are approximate (rounded). Numbers in parentheses indicate the number of lines excluding blanks and comments. OS: Open source. CC: Creative Commons. CPC: Computer Physics Communications. GPL: GNU General Public License. MIT: Massachusetts Institute of Technology. LBM: Lattice Boltzmann Method.} 
\label{tab:1}
\setlength{\tabcolsep}{0.4em}
\begin{tabular}{@{}p{0.7cm}lp{4.4cm}p{4.3cm}p{3.5cm}lp{0.975cm}p{0.625cm}@{}}
\toprule
\textbf{Year} & \textbf{Name} & \textbf{Model description} & \textbf{Implementation} & \textbf{Performance} & \textbf{OS} & \textbf{License} & \textbf{Ref.}\\
\midrule
1990 & 2D-FROTH & dry foams in mechanical equilibrium, shared polygonal boundaries & 8200 (4500) lines of Fortran 77 code, three library dependencies & demonstrated simulations with a few dozen bubbles & yes & CPC & \cite{Kermode:1990}\\
\midrule
1992-- 1996 & PLAT & wet foams in mechanical equilibrium, free polygonal boundaries & \num{9800} (5100) lines of C code, requires X/Motif libraries for GUI & demonstrated simulations with a few dozen bubbles & yes & public domain & \cite{Bolton:1992,PLAT:1996}\\
\midrule
2005 & --- & tumor growth, free cell boundaries coupled to a Navier--Stokes solver & Fortran & demonstrated tumor tissue growth to about 900 cells & no & --- & \cite{Rejniak:2005}\\
\midrule
2008 & --- & meshed elastic cell walls immersed in a Navier--Stokes fluid & --- & demonstrated growth to about 800 cells & no & --- & \cite{Dillon:2008}\\
\midrule
2010 & --- & viscoelastic polygonal cells with cytoskeletal elements & --- & demonstrated simulations with a few hundred cells & no & --- & \cite{Jamali:2010}\\
\midrule
2011 & --- & gastrulation with elastic polygonal cell boundaries & in C using CUDA for parallelization on the GPU; reimplemented in Python using Numba \cite{Sande:2020} & demonstrated simulations with a few dozen cells & no & --- & \cite{Tamulonis:2011}\\
\midrule
2011 & VirtualLeaf & framework for plant growth with shared polygonal cell walls & \num{22000} (\num{14000}) lines of C++ code, GUI based on Qt 4.6, dependencies on libiconv, libxml2, libz; no longer maintained & demonstrated simulations with a few hundred cells & yes & GPL v2 & \cite{Merks:2011}\\
\midrule
2014 & DySMaL & deformable bubble model for wet foams & Fortran program parallelized with MPI and OpenMP & 50 million timesteps with 1700 bubbles took 5 hours on 16 cores & no & --- & \cite{Kahara:2014}\\
\midrule
2014 & EpiCell2D & mechanical model for tissue growth with free cell boundaries & Fortran program, parallelized with MPI & dynamic simulations with a few hundred cells \cite{Mkrtchyan:2014, Madhikar:2021} & no & --- & \cite{Mkrtchyan:2014}\\
\midrule
2015 & LBIBCell & polygonal cell boundaries immersed in a LBM fluid with chemical signaling & \num{25000} (\num{13000}) lines of C++ code, parallelized with OpenMP, dependencies on Boost, VTK \& CMake; no longer maintained & can grow a tissue to \num{10000} cells in a day & yes & MIT & \cite{Tanaka:2015}\\
\midrule
2017 & --- & polygonal cell boundaries immersed in viscous Newtonian fluid & built on Chaste/cell\_based \cite{Pitt:2009} which depends on many third-party packages, \num{130000} (\num{63000}) lines of C++ code in total & 2000 timesteps with 20 cells in a few minutes on 6 cores & yes & GPL v3 & \cite{Cooper:2017}\\
\midrule
2018 & DP model & packing of elastic particles with free polygonal boundaries & --- & demonstrated simulations with up to 1000 particles & no & --- & \cite{Boromand:2018}\\
\midrule
2016-- 2022 & NCC model & cell migration, polygonal boundaries with protrusions, coupled to reaction-diffusion solver & \num{19000} (\num{12000}) lines in Python using Numba \cite{numba-ncc:2016}; reimplemented with 9600 (7600) lines in Rust \cite{rust-ncc:2020} & single-threaded runtime of 6 hours for 49 cells simulated over 10 hours & yes & MIT or Apache & \cite{Merchant:2018}\\
\midrule
2021 & LBfoam & dry and wet foams based on the LBM, similar to earlier unnamed closed-source program \cite{Korner:2002} & \num{350000} (\num{250000}) lines of C++ code, including Palabos \cite{Latt:2020} & demonstrated HPC simulations with up to 300 gas cells, parallelized with MPI & yes & AGPL v3 & \cite{Ataei:2021}\\
\midrule
2021 & PalaCell2D & polygonal cell boundaries immersed in a LBM fluid with chemical signaling & \num{11000} (8600) lines of C++ code building on Palabos, dependencies on TinyXML-2 and CMake & demonstrated simulations with up to 400 cells & no & --- & \cite{Conradin:2021}\\
\midrule
2021 & --- & shared fluctuating polygonal cell boundaries, intercellular spaces & 2900 (2300) lines of Matlab code & demonstrated simulations with a few dozen cells & yes & public domain & \cite{Kim:2021}\\
\midrule
2021 & EdgeBased & suite of cell-based tissue models, one with polygonal cell boundaries & \num{14000} (6800) lines of Matlab code & tissue growth to 600 cells with 10 vertices each in 10\,h & yes & GPL v3 & \cite{Brown:2021}\\
\midrule
2023 & Epimech & epithelial monolayer tissues mechanically coupled to a substrate & \num{25000} (\num{13000}) lines of Matlab code, with a GUI & growth to 1600 cells in 20 hours without substrate & yes & GPL v3 & \cite{Tervonen:2023}\\
\bottomrule
\end{tabular}
\end{table*}

While most of these computer programs were developed for a specific biological or physical application, quantifications of their computational performance and scalability are rare. Where it can be estimated from communicated cell counts, the used parallelization strategy or approximate runtimes, typical simulations with some dozens to a few thousand cells or bubbles require in the order of minutes to days of wallclock time. One notable exception appears to be \textit{DySMal} with a large number of timesteps performed in comparably short time, although the paper did not show simulations with more than 1700 bubbles \cite{Kahara:2014}.

There are also a number of phase-field models to simulate the fluid dynamics of cell monolayers and foams \cite{Nonomura:2012, Palmieri:2015, Lober:2015, Biner:2017, Jiang:2019, Lavoratti:2021, Lecrivain:2021}. The numerical burden of these grid-based approaches is large, though. Reports of the computational performance of these programs, where disclosed, range from displayed simulations of 12 cells on 36 processors \cite{Palmieri:2015} to simulations with 100 cells that require a month of runtime on 16 cores \cite{Lavoratti:2021, Lecrivain:2021}. Larger systems appear to be out of reach for these programs, possibly with the exception of the first phase-field model of biological cells \cite{Nonomura:2012}, which was implemented in Fortran, but is not publicly available.

An additional challenge faced with most of the published programs is their level of code complexity. While the leanest implementations comprise a few thousand lines of code \cite{Kermode:1990, PLAT:1996, Kim:2021}, the majority are found in the tens to hundreds of thousand lines \cite{Merks:2011, Tanaka:2015, Cooper:2017, Ataei:2021, Tervonen:2023}. This is sometimes resulting from larger frameworks they are built on or into \cite{Cooper:2017, Ataei:2021, Conradin:2021}, and can form an obstacle for their usability to other researchers, and make it harder to maintain the code.

Aside from these technical aspects, there are also methodological improvements needed to enhance the range of phenomena that can be simulated with such models. Emulsions, but also wet foams and a variety of developing biological tissues can undergo a wide range of structural changes that alter the topology at the level of bubbles or cells: they can merge or split up, aggregate or segregate, engulf or expel each other, and emerge or disintegrate. In a biological context, cells vanish from a tissue for example through apoptosis or extrusion, and they divide through mitosis. Cell fusion, on the other hand, occurs in failed cytokinesis \cite{Jantsch:1999}, tumor progression \cite{Lu:2009}, myoblasts \cite{Rochlin:2010}, and transport vesicles \cite{Stillwell:2016}. With the existing computer programs, such fusion processes cannot readily be simulated.

With this paper, we introduce \program{}, a portable, compact and lightweight C++ program that overcomes these limitations. \program{} is a portmanteau of polygon and hoop, with intentional ambiguity in the meaning of ``poly'', hinting at the fact that our program is designed to represent large systems consisting of many hoops or polygons. Compared to similar programs, \program{} offers a reduction in code volume by an order of magnitude or more, and a speedup of several orders of magnitude in cases where no explicit fluid coupling is needed, all while providing an extended feature set including the above-mentioned topological changes, in a standalone program. Comprising about 720 lines of commented code, \program{} is designed to be as compact and simple as possible, allowing also novices to follow the implementation, modify it, and run simulations. \program{} is devoid of magical numbers, error tolerances, and iterative solvers. With only 23 geometrical and physical parameters, a large variety of different dynamical systems can be simulated. 

Drawing inspiration from previous models \cite{Kahara:2014, Tanaka:2015, Boromand:2018, Conradin:2021}, \program{} represents curved hoops with high spatial resolution, and allows for interstitial volume through separate representation of adjacent cell boundaries, unlike vertex models \cite{Kawasaki:1989, Nagai:2001, Farhadifar:2007, Hufnagel:2007, Kim:2021}. It offers the topological freedom to simulate spontaneous engulfment, splitting, fusion etc., controlled by a minimal set of parameters. System sizes in the order of a million hoops can be simulated on conventional computers, and we report the serial and parallel computational performance in easy-to-reproduce benchmarks. Combining computational performance with spontaneous topological changes, we expect \program{} to bridge the gap between microscopic events such as cell fusion, and macroscopic tissue structure or function, in future computational research.

\section{Physical model}

\subsection{Continuum description}

\begin{figure}
	\centering
	\includegraphics[width=\linewidth]{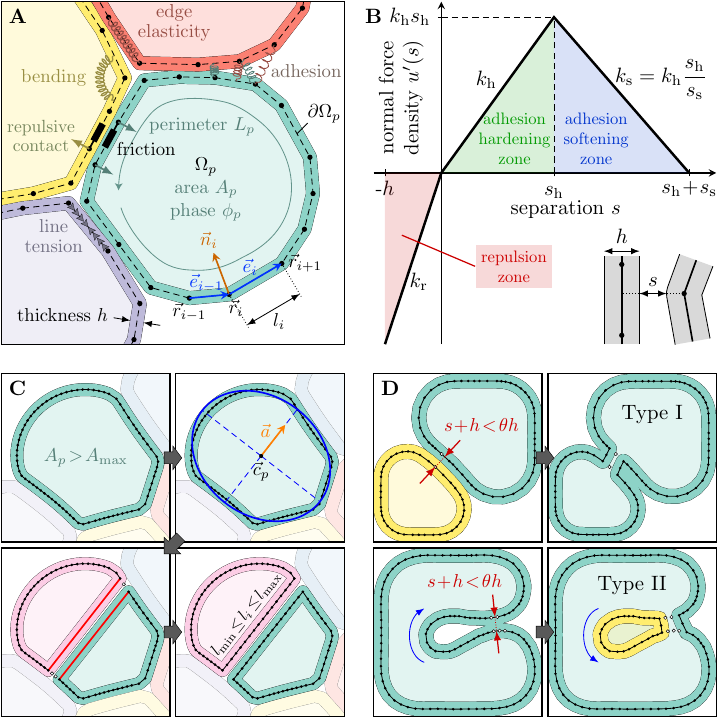}
	\caption{\textbf{Two-dimensional soft polygon model.}
	\textbf{A}, Model overview. Particles or cells are represented by interacting polygons (colored) with edge thickness $h$.
	\textbf{B}, Trilinear normal contact force model, governed by four material parameters $k_\mathrm{r}$, $k_\mathrm{h}$, $s_\mathrm{h}$, $s_\mathrm{s}$. Positive forces represent attraction, modeled by a classical bilinear traction-separation law (slopes $k_\mathrm{h}$ and $-k_\mathrm{s}$). $s=0$ corresponds to two segments just touching.
	\textbf{C}, Cell division model. Cells whose area exceeds a division threshold are cut in half in direction of the minor axis ($\vec{a}$, orange) of their inertia ellipse (blue), removing too close vertices (white). The two new edges (red) are then refined to restore the desired mesh resolution.
	\textbf{D}, Fusion model. Polygon pairs whose (negative) separation $s$ subceeds a fusion threshold $\theta h-h$ are merged by breaking up and rejoining their edges. White vertices are removed. Two fusion types are modeled: I: merger of two touching polygons (top); II: split of a self-touching polygon into two (bottom). Both types can involve external or internal (enclosed) polygons. Internalized or externalized polygons are reoriented into anti-clockwise (blue arrows).
	}
	\label{fig:1}
\end{figure}

\program{} simulates the Newtonian dynamics of ensembles of $N$ closed hoops representing the boundaries of interacting fluidic particles (biological cells, gas bubbles, fluid droplets etc.). The interior $\Omega_p$ of particle $p$ is not explicitly modeled; instead, we parameterize the position of its boundary with the vector field $\vec{r}(l)=[x(l),y(l)]^\top\in\partial\Omega_p$ (Fig.~\ref{fig:1}A), and describe the ensemble by the potential energy
\begin{equation}
\label{eq:U}
\begin{split}
U = \sum_{p=1}^N \bigg(&\frac{k_\mathrm{a}}{2}\left(A_p-A_{p,0}\right)^2 + \gamma L_p + \int_0^{L_{p,0}}\!\!\left(\frac{k_\mathrm{l}}{2}\varepsilon^2+\frac{k_\mathrm{b}}{2}\kappa^2\right)dl\\
&+\rho g \phi_p S_{x,p} + \rho_\mathrm{l} g_\mathrm{l} \oint_{\partial\Omega_p}\!y\,dl + \sum_{q=1}^NU_\mathrm{int}(p,q)\bigg).
\end{split}
\end{equation}
The first of the six terms in $U$ is a linearized form of elastic compression with a 2D bulk modulus of the enclosed medium of $k_\mathrm{a}A_{p,0}$, penalizing deviations of the current particle area $A_p$ from a target area $A_{p,0}$. The second term represents line tension with strength $\gamma$ and hoop length $L_p$. In the third summand, we account for linearly elastic tension, compression and bending of each hoop by integrating over their reference contours with length $L_{p,0}$. In the integrand, $\varepsilon=\norm{d\vec{r}/dl}-1$ is the local Cauchy strain, $\kappa=\norm{d^2\vec{r}/dl^2}$ the local curvature, $l\in[0,L_{p,0}]$ the arclength parameter running along the hoop's contour, $k_\mathrm{l}$ and $k_\mathrm{b}$ the elastic dilatation and bending moduli.

To model floppy elastic particles, we couple their target areas and perimeters through the isoparametric ratio (or ``asphericity'' \cite{Boromand:2018})
\begin{equation}
Q = \frac{L_{p,0}^2}{4\pi A_{p,0}}
\end{equation}
which equals $1$ for an unstrained circle. This coupling is unidirectional, defining the hoop length $L_{p,0}$ from a given area $A_{p,0}$.

With the fourth energy term in Eq.~\ref{eq:U}, we model hydrostatic pressure, relevant primarily for the simulation of vertical bubbly liquids and emulsions. $g$ is the gravitational acceleration, $\rho$ the 2D mass density difference between the inner and outer media, and $\phi_p=\pm1$ a binary flag indicating the ``phase'' enclosed by hoop $p$. In a biological context where the particles represent cells, $\phi_p$ may for example be used to discriminate the cell cytosol from intra- or extracellular components, such as organelles, the extracellular matrix, etc. In simulations of a binary immiscible fluid, it discriminates between bubbles and the medium. The sign convention is such that a submerged particle $p$ containing phase $\phi_p=-1$ is buoyant, if $\rho, g>0$. $S_{x,p}$ is the particle's first moment of area (defined in Eq.~\ref{eq:Sx}). With the fifth energy term, we allow for a complementary representation of gravity, acting on the particle boundary $\partial\Omega_p$ instead of its bulk $\Omega_p$. $g_\mathrm{l}$ is the gravitational acceleration for the hoop contour, and $\rho_\mathrm{l}$ its line mass density (mass per unit contour length). The sixth and final energy term in Eq.~\ref{eq:U} is a double sum running over all particle pairs $(p,q)$, accounting for bilateral interactions with an interaction potential
\begin{equation}
\label{eq:Uint}
U_\mathrm{int}(p,q) = \rho_\mathrm{l}^2\oint_{\partial\Omega_p} \oint_{\partial\Omega_q} u\big(s(l_q,l_p)\big)\,dl_q\,dl_p,
\end{equation}
where $s(l_q,l_p)=\norm{\vec{r}(l_p)-\vec{r}(l_q)}-h$ is the spatial separation between the hoops $p$ and $q$ at their arclength positions $l_p$ and $l_q$, and $h>0$ is the (virtual or physical) hoop thickness (Fig.~\ref{fig:1}A). The interaction density $u(s)$ implemented in \program{} models steric repulsion and adhesion in the perhaps simplest possible way, such that a piece-wise linear traction-separation law results from it, divided into three zones (Fig.~\ref{fig:1}B). Upon volumetric overlap (negative separation $s$), hoop segments repel each other with repulsion strength $k_\mathrm{r}$. If the separation is positive, hoop segments attract each other with adhesion strength $k_\mathrm{h}$ up to a maximum separation $s_\mathrm{h}$, beyond which the adhesion softens all the way to zero again at $s=s_\mathrm{h}+s_\mathrm{s}$ to ensure a continuous force transmission. Formally, this can be expressed by
\begin{equation}
u(s) = \frac{1}{2}
\begin{cases}
k_\mathrm{r}s^2 & s \leq 0\\
k_\mathrm{h}s^2 & 0 \leq s \leq s_\mathrm{h}\\
k_\mathrm{h}s_\mathrm{h}^2 + k_\mathrm{s}(s-s_\mathrm{h})(2s_\mathrm{s}+s_\mathrm{h}-s) & s_\mathrm{h} \leq s \leq s_\mathrm{h} + s_\mathrm{s}\\
k_\mathrm{h}s_\mathrm{h}^2 + k_\mathrm{s}s_\mathrm{s}^2 & s_\mathrm{h} + s_\mathrm{s} \leq s
\end{cases}
\end{equation}
where $k_\mathrm{s}=k_\mathrm{h}s_\mathrm{h}/s_\mathrm{s}$ is the adhesion softening strength. The resulting normal force density is then given by the derivative w.r.t.\ the separation,
\begin{equation}
u'(s) = 
\begin{cases}
k_\mathrm{r}s & s \leq 0\\
k_\mathrm{h}s & 0 \leq s \leq s_\mathrm{h}\\
k_\mathrm{s}(s_\mathrm{h}+s_\mathrm{s}-s) & s_\mathrm{h} \leq s \leq s_\mathrm{h} + s_\mathrm{s}\\
0 & s_\mathrm{h} + s_\mathrm{s} \leq s
\end{cases}.
\end{equation}

Note that this model allows for self-interactions ($p=q$), which are relevant especially for floppy particles (large $Q$ or small $k_\mathrm{a}$). To exclude false detection of steric repulsion between nearby segments of the same hoop, we set $u\big(s(l_q,l_p)\big)=0$ if both $p=q$ and the arclength distance between the contacting points is too short, i.e., $\norm{l_p-l_q}\leq h$.

\subsection{Discretization as polygons}

We discretize the particle boundaries as polygons that automatically remodel if needed to maintain a uniform spatial resolution. Each polygon consists of a list of vertex positions $\vec{r}_i=[x_i,y_i]^\top$, $i=1,...,M_{p}$. The particle area is calculated with the shoelace formula, a special case of Green's theorem:
\begin{equation}
A_p = \iint_{\Omega_p}\!\!dx\,dy = \frac{1}{2}\abs{\oint_{\partial\Omega_p}\!\!x\,dy-y\,dx} \approx \frac{1}{2}\abs{\sum_{i=1}^{M_p}\vec{r}_i \wedge \vec{r}_{i+1}}
\end{equation}
where
\begin{equation}
\vec{r}_i \wedge \vec{r}_j = x_i y_j-x_j y_i
\end{equation}
is the 2D exterior vector product. The vertex list is cyclic, i.e., $\vec{r}_{M_p+1}=\vec{r}_1$. For vertices arranged in counter-clockwise orientation, as implemented here, the absolute value bars can be dropped. From the polygon edges $\vec{e}_i = \vec{r}_{i+1}-\vec{r}_i$ (Fig.~\ref{fig:1}A), the hoop length can be approximated as
\begin{equation}
L_p = \oint_{\partial\Omega_p}\!\!\norm{d\vec{r}} \approx \sum_{i=1}^{M_p}l_i
\end{equation}
where
\begin{equation}
l_i = \norm{\vec{e}_i} = \sqrt{(x_{i+1}-x_i)^2+(y_{i+1}-y_i)^2}.
\end{equation}
For the elastic line integral in Eq.~\ref{eq:U}, we sum up the squared edge strains $\varepsilon_i = l_i/l_{i,0}-1$, weighted by the reference edge lengths $l_{i,0}$:
\begin{equation}
\int_0^{L_{p,0}}\!\!\varepsilon^2\,dl \approx \sum_{i=1}^{M_p}\varepsilon_i^2\,l_{i,0}
\end{equation}
For the bending energy, we follow the discretization proposed in \cite{Bergou:2008}:
\begin{equation}
\int_0^{L_{p,0}}\!\!\kappa^2\,dl \approx \sum_{i=1}^{M_p}\kappa_i^2\,\overline{l}_{i,0}
\end{equation}
with nodal curvature
\begin{equation}
\kappa_i = \frac{2\vec{e}_{i-1} \wedge \vec{e}_i}{\overline{l}_{i,0}(l_{i-1,0}l_{i,0}+\vec{e}_{i-1}\cdot\vec{e}_i)}
\end{equation}
and an average reference length of the edges incident on node $i$,
\begin{equation}
\overline{l}_{i,0} = \frac{l_{i-1,0}+l_{i,0}}{2}.
\end{equation}

To discretize the gravitational potential acting on the particle area, we require the first moment of area of each polygon, $S_{x,p}$. It can be computed as \cite{Soerjadi:1968}
\begin{equation}
\label{eq:Sx}
S_{x,p}=\iint_{\Omega_p}\!y\,dx\,dy \approx \frac{1}{6}\sum_{i=1}^{M_p}(\vec{r}_i \wedge \vec{r}_{i+1}) (y_i+y_{i+1}).
\end{equation}
To discretize the boundary gravitational potential, we introduce the nodal mass $m_i=\rho_\mathrm{l}\overline{l}_i$. As we will later dynamically adapt the polygonal discretization to keep edge lengths in a predefined range, these nodal masses do not vary greatly. Depending on the use case, and in particular with fluidic interfaces in mind, where the polygon vertices represent the interfacial shape but carry no mass, small differences in nodal inertia may be irrelevant or even undesired. We therefore harmonize the vertex masses to a constant value $m$, and will later remove this mass scale entirely by expressing all massive parameters relative to it. With this simplification, the contour mass density $\rho_\mathrm{l}$ is eliminated from the model and the boundary gravitational potential becomes
\begin{equation}
\rho_\mathrm{l} g_\mathrm{l} \oint_{\partial\Omega_p}\!y\,dl \approx \rho_\mathrm{l} g_\mathrm{l} \sum_{i=1}^{M_p}y_i\overline{l}_i \approx g_\mathrm{l} m\sum_{i=1}^{M_p}y_i
\end{equation}
such that the effect of it on all polygon vertices is a constant uniform downward acceleration $g_\mathrm{l}$.

Finally we also discretize the interaction potential. Eq.~\ref{eq:Uint} becomes a double sum over all pairs of vertices,
\begin{equation}
\label{eq:Uint_discrete}
U_\mathrm{int}(p,q) \approx m^2\sum_{i=1}^{M_p} \sum_{j=1}^{M_q} u(s_{ij})
\end{equation}
subject to the condition that the interacting vertex pair $i,j$ is either on different polygons ($p\neq q$), or else, further than $h$ apart along the polygon. $s_{ij}$ denotes the separation between vertex $i$ on polygon $p$ and its closest point of approach on the two edges incident on vertex $j$ on polygon $q$. Note that this definition is asymmetric, hence the full symmetric double sum in Eq.~\ref{eq:Uint_discrete}. Formally, this can be expressed as
\begin{equation}
\label{eq:sij}
s_{ij} = \min_{\xi\in[-1,1]}\norm{\vec{r}_j(\xi)-\vec{r}_i} - h
\end{equation}
where
\begin{equation}
\vec{r}_j(\xi) = \begin{cases}
\vec{r}_j+\xi(\vec{r}_{j+1}-\vec{r}_j) & \xi \geq 0\\
\vec{r}_j+\xi(\vec{r}_{j-1}-\vec{r}_j) & \xi \leq 0\\
\end{cases}
\end{equation}
is the closest point on the two edges next to vertex $j$. The barycentric edge coordinate of the closest point of approach, $\xi$, allows the resulting interaction forces to be distributed to the involved vertices in proportion.

\subsection{Vertex forces}

From the discretized potential, the conservative nodal forces can be derived. The gradient w.r.t.\ the degrees of freedom of vertex $i$ of polygon $p$ reads
\begin{equation}
\label{eq:gradU}
\begin{aligned}
\vec{\nabla}_i U =& \begin{bmatrix}
\partial U/\partial x_i\\
\partial U/\partial y_i
\end{bmatrix}\\
\approx& -\frac{k_\mathrm{a}}{2} (A_p-A_{p,0})\vec{n}_i + \gamma \left(\vec{t}_{i-1} - \vec{t}_i\right) + k_\mathrm{l} \left(\varepsilon_{i-1}\vec{t}_{i-1}-\varepsilon_i\vec{t}_i\right)\\
&+ 4k_\mathrm{b}\left(a_{i-1}\frac{\vec{e}_{i-2}^\perp - a_{i-1}\vec{e}_{i-2}}{\overline{l}_{i-1}b_{i-1}}\right. - a_i\frac{\vec{n}_i+a_i(\vec{e}_i-\vec{e}_{i-1})}{\overline{l}_i b_i}\\
&\hspace{2.4em}+ \left.a_{i+1}\frac{\vec{e}_{i+1}^\perp + a_{i+1}\vec{e}_{i+1}}{\overline{l}_{i+1}b_{i+1}}\right)\\
&-\frac{\rho g \phi_p}{6}\left((y_{i-1}+y_i+y_{i+1})\vec{n}_i - \begin{bmatrix}
0\\
d_i
\end{bmatrix}\right) + \begin{bmatrix}
0\\
m g_\mathrm{l}
\end{bmatrix}\\
&+ m^2\sum_{p=1}^N\sum_{q=1}^N\sum_{i=1}^{M_p} \sum_{j=1}^{M_q} u'(s_{ij})\vec{\nabla}_i s_{ij}.
\end{aligned}
\end{equation}
Here, we used the following notation:
\begin{equation}
\label{eq:ni}
\vec{n}_i = (\vec{e}_{i-1}+\vec{e}_i)^\perp
\end{equation}
is the unnormalized inward normal vector at vertex $i$ (Fig.~\ref{fig:1}A),
\begin{equation}
\vec{t}_i = \frac{\vec{e}_i}{l_i}
\end{equation}
the unit tangent vector (director) of the edge following vertex $i$, and
\begin{equation}
a_i = \frac{d_i}{b_i},\qquad d_i = \vec{e}_{i-1} \wedge \vec{e}_i,\qquad b_i = l_{i-1}l_i + \vec{e}_{i-1}\cdot\vec{e}_i.
\end{equation}
With the $^\perp$ symbol we denote the perpendicular vector:
\begin{equation}
\begin{bmatrix}x\\y\end{bmatrix}^\perp = \begin{bmatrix}-y\\x\end{bmatrix}.
\end{equation}
Notice that in Eq.~\ref{eq:gradU}, we assume inextensibility of the polygon edges in the bending forces for simplicity, as proposed in \cite{Bergou:2008}. Moreover, to avoid the discontinuity in the coordinate of the closest point of approach between interacting hoop edges \cite{Vetter:2013}, we set $\vec{\nabla}_i\xi=0$, such that the contact forces are applied in normal direction, which considerably simplifies the interaction expression:
\begin{equation}
-\vec{\nabla}_i s_{ij} \approx \vec{n}_{ij}=\frac{\vec{r}_j(\xi)-\vec{r}_i}{\norm{\vec{r}_j(\xi)-\vec{r}_i}}.
\end{equation}
Finally, we note that the quadruple sum in the interaction term in Eq.~\ref{eq:gradU} is sparse and need not actually be evaluated as such, because the interaction forces are local ($u'(s_{ij})=0$ for $s_{ij}\geq s_\mathrm{h}+s_\mathrm{s}$). Instead, spatial partitioning can be used to find non-zero contributions efficiently (see Sec.~\ref{sec:contact}).

With the gradient of the potential defined, we can express the vertex forces. The total force vector acting on vertex $i$ is the sum of conservative and dissipative forces:
\begin{equation}
\vec{f}_i = -\vec{\nabla}_i U - c_\mathrm{v}\dot{\vec{r}}_i - \frac{\rho c_\mathrm{d}}{4}\abs{\dot{\vec{r}}_i\cdot\vec{n}_i}\dot{\vec{r}}_i - \sum_{i,j}{}^{'}\!\!\left( c_\mathrm{c}\vec{v}_\perp + \mu\norm{\vec{f}_\perp}\frac{\vec{v}_\parallel}{\norm{\vec{v}_\parallel}}\right).
\end{equation}
Here, the first force term combines all conservative forces derived from the potential, including steric repulsion and adhesion. The second term models viscous damping with a global coefficient $c_\mathrm{v}$. As a second mode of energy dissipation, we include drag (third force term), which is proportional to the squared vertex velocity, as well as to the edge length, via Eq.~\ref{eq:ni}. $c_\mathrm{d}$ is the dimensionless drag coefficient, and $\rho$ the mass density difference between the inner and outer phases as used in Eq.~\ref{eq:U}. In the fourth term, we sum over all global pairs of interacting vertices $(i,j)$ and add, for each interaction, a collision damping term in normal direction with coefficient $c_\mathrm{c}$, and a frictional force in tangential direction with dynamic Coulomb friction coefficient $\mu$, proportional to the modulus of the normal interaction force 
\begin{equation}
\vec{f}_\perp = m^2 u'(s_{ij})\vec{n}_{ij} - c_\mathrm{c}\vec{v}_\perp.
\end{equation}
To arrive at this decomposition, we split the relative velocity
\begin{equation}
\vec{v}_{ij} = \dot{\vec{r}}_j(\xi) - \dot{\vec{r}}_i
\end{equation}
(with $\xi$ from Eq.~\ref{eq:sij}) into a perpendicular and a parallel part:
\begin{equation}
\vec{v}_\perp = \left(\vec{v}_{ij}\cdot\vec{n}_{ij}\right)\vec{n}_{ij},\qquad\vec{v}_\parallel = \vec{v}_{ij} - \vec{v}_\perp.
\end{equation}
Friction is only added if $\norm{\vec{v}_\parallel}>0$. Finally, using the total forces, Newton's second law is solved for each polygon vertex:
\begin{equation}
m \ddot{\vec{r}}_i = \vec{f}_i.
\end{equation}

\section{Numerical implementation}

\subsection{Remodeling}

\program{} is designed to enable arbitrarily large deformations. For applications in fluid dynamics or tissue biology, where the hoops represent fluidic interfaces, lipid bilayers, etc., dynamic remodeling of the particle/cell boundaries is essential to maintain good quality in the polygonal discretization. The polygons are automatically remodeled in each timestep to maintain an approximately uniform spatial discretization. Edges whose length exceeds a maximum value, $l_i>l_\mathrm{max}$, are bisected, introducing a new vertex in the middle. Edges whose length subceeds a minimum value, $l_i<l_\mathrm{min}$, are removed by merging their two vertices into one, positioned at the edge midpoint. Edge lengths thus remain in a predefined range, $l_i\in[l_\mathrm{min},l_\mathrm{max}]$. These refining and coarsening steps intentionally break mass and momentum conservation of the particle boundaries $\partial\Omega_p$ to enable massive growth, as vertices with mass $m$ are added or removed. They do, however, conserve the masses of the enclosed particles $\Omega_p$ themselves, as their target areas $A_{p,0}$ are unaffected.

\subsection{Polygon growth, removal and division}

\program{} can grow or shrink polygons over time by changing their target area according to
\begin{equation}
\frac{dA_{p,0}}{dt}=\alpha
\end{equation}
where $\alpha$ is a constant area growth rate. For example for tissue simulations with biological variability, $\alpha$ can be drawn from a random distribution for each cell individually.

Polygons whose area drops below a specified minimal value, $A_p<A_\mathrm{min}$, are removed. This feature is useful in particular for the simulation of cell extrusion from epithelial monolayers, through apoptosis, active contraction, or overcrowding.

Primarily for biological applications of proliferative tissues, polygons (cells) are divided into two when their area exceeds a threshold value, $A_p>A_\mathrm{max}$ (Fig.~\ref{fig:1}C). Like the growth rate, the maximum cell area $A_\mathrm{max}$ can be cell-specific and drawn from a random distribution to introduce cell-to-cell variability. Division is set to occur always through the centroid \cite{Soerjadi:1968}
\begin{equation}
\vec{c}_p = \begin{bmatrix}c_x\\c_y\end{bmatrix} = \frac{1}{A_p}\iint_{\Omega_p}\!\begin{bmatrix}x\\y\end{bmatrix}dx\,dy = \frac{1}{6A_p}\sum_{i=1}^{M_p} (\vec{r}_i \wedge \vec{r}_{i+1}) (\vec{r}_i+\vec{r}_{i+1}).
\end{equation}
The long axis rule is implemented, according to which cells divide in direction of their longest extent. To find the division axis $\vec{a}$ perpendicular to it (Fig.~\ref{fig:1}C, orange), we use the eigensystem of the cell's inertia tensor, which is an intuitive way to define a cell's orientation in space by approximating it by its own inertia ellipse (Fig.~\ref{fig:1}C, blue). For a polygon with uniform unit mass density, the inertia tensor reads \cite{Soerjadi:1968}
\begin{equation}
\mathbf{J}_p =\frac{1}{12}\sum_{i=1}^{M_p} (\vec{r}_i \wedge \vec{r}_{i+1}) \begin{bmatrix}i_{xx} & i_{xy}\\i_{xy} & i_{yy}\end{bmatrix}
\end{equation}
where
\begin{equation}
\begin{split}
i_{xx} &= y_i^2 + y_i y_{i+1} + y_{i+1}^2\\
i_{yy} &= x_i^2 + x_i x_{i+1} + x_{i+1}^2\\
i_{xy} &= x_i y_i + x_{i+1} y_{i+1} + (x_i y_{i+1} + x_{i+1} y_i) / 2
\end{split}.
\end{equation}
The inertia tensor about the polygon centroid can be computed by applying the parallel axis theorem, resulting in
\begin{equation}
\mathbf{I}_p = \begin{bmatrix} I_{xx} & I_{xy}\\I_{xy} & I_{yy}\end{bmatrix} = \mathbf{J}_p - A_p \begin{bmatrix}c_y^2 & -c_x c_y\\ -c_x c_y & c_x^2\end{bmatrix}
\end{equation}
$\mathbf{I}_p$ is symmetric positive semi-definite, its eigenvalues are the principal moments of inertia, and its eigenvectors are the principal axes. The shortest axis is the eigenvector with the largest eigenvalue $\lambda$. An efficient and robust way to find this (unnormalized) axis $\vec{a}$ numerically is \cite{Knill:2004}
\begin{equation}
\vec{a} = \begin{cases}
\begin{bmatrix}I_{xy}\\\lambda - I_{xx}\end{bmatrix} & \text{if } I_{xx} < I_{yy}\\
\begin{bmatrix}\lambda - I_{yy}\\I_{xy}\end{bmatrix} & \text{else}
\end{cases}
\end{equation}
with
\begin{equation}
\lambda = \Delta I + I_{yy} + \sqrt{\Delta I^2 + I_{xy}^2}, \qquad \Delta I=\frac{I_{xx} - I_{yy}}{2}.
\end{equation}
Note that for a perfectly rotationally symmetric polygon, $I_{xx}=I_{yy}$ and $I_{xy}=0$, resulting in $\vec{a}=\vec{0}$. In this special case, a random division axis is drawn instead.

With the line of division defined, the polygon is cut into two at its points of intersection with the division line, and vertices are iteratively removed if necessary to prevent overlaps between the two newborn daughters (Fig.~\ref{fig:4}C, white dots). The polygons are then resealed with straight lines (strained equally to the remainder of the polygon), and the edge remodeling algorithm outlined above restores the desired spatial resolution (Fig.~\ref{fig:4}C, bottom row). The mother cell's target area is distributed to its children in proportion to their actual area, and both children finally draw new (random) area growth rates $\alpha$ and division areas $A_\mathrm{max}$.

\subsection{Polygon fusion}

Topological changes are the most computationally demanding and programmatically complex part of \program{}, making up more than a third of the code and accounting for about half of the total runtime in a typical simulation when enabled. Aside from division and removal described above, two further types of topological changes (I and II) are implemented, which increment or decrement the number of polygons in the system, subsumed under ``polygon fusion'' here (Fig.~\ref{fig:1}D). Type I refers to the merger of two touching distinct polygons into one, which decrements the polygon count. Type II refers to ``self-fusion'', i.e., the splitting of a polygon into two due to two distant segments of the same polygon touching, which increments the polygon count. Both fusion types can occur in two variants each: For Type I, the two touching polygons may either be external to each other (as shown in Fig.~\ref{fig:1}D, top row) or one inside the other. Conversely, for Type II, the newly spawned polygon may either be outside the existing one or inside (as shown in Fig.~\ref{fig:1}D, bottom row). When polygons internalize or externalize in this process, they are reoriented into anti-clockwise vertex order, as \program{} requires all polygons to be anti-clockwise for simplicity. Whether a polygon is enclosing the other is tested with an efficient ray casting algorithm \cite{Franklin:2006} during the fusion event.

Fusion events are triggered based on the local degree of mutual interpenetration of pairs of polygon segments, which is equivalent to contact stress criteria that depend on the mutual contact depth, such as in Hertzian models. We opted for a minimal, generic model with a single, intuitive and easy-to-control parameter, the fusion threshold $\theta\in[0,1]$. If $\theta=0$, the fusion feature is disabled. If $\theta=1$, polygons fuse immediately when they come in physical contact. In general, a value in between lets polygons fuse when they press against one another sufficiently to let $s+h<\theta h$, with a negative separation $s$ between polygon segments (Fig.~\ref{fig:1}B). When this overlap criterion is met, the polygons are broken up and their vertices are iteratively removed in the vicinity of the closest point of contact, until they lie sufficiently (i.e., a minimal distance of $h$) apart such that the two gaps can be rejoined without residual overlaps (Fig.~\ref{fig:1}D).

\subsection{Contact detection}
\label{sec:contact}

Hoop interactions, if enabled, can make up a large fraction of the overall computational cost, because generally, any pair of vertices can be in contact. To reduce the quadruple sum in Eq.~\ref{eq:gradU} to effectively a double sum running over all polygons and their vertices, we employ spatial partitioning \cite{Quentrec:1973} using linked lists. Within the global bounding box of all polygons, the simulation space is divided into square boxes with side length $\max_i\{l_i\} + h + s_\mathrm{h} + s_\mathrm{s}$, which ensures that interacting vertex pairs are no more than one box apart. Instead of the current largest edge length $\max_i\{l_i\}$, the upper bound $l_\mathrm{max}$ could also be used for simplicity, but we observed generally better performance with the former. A global array then stores a pointer to the first vertex in each box, and each vertex holds a pointer to the next vertex in the same box. Checking for polygon interactions then reduces to a loop over all vertices $i$ and a loop over all vertices contained in the local group of $3\times3$ boxes around $i$. This procedure reduces the time complexity of contact detection from squared to linear in the number of vertices. The boxes are recomputed in each timestep for simplicity.

\subsection{Time integration}

For the discontinuous evolution of a particle system with instantaneous topological changes as implemented in \program{}, multi-step or implicit time integration methods are difficult to formulate. We therefore propagate the vertex positions $\vec{r}_i$ and velocities $\vec{v}_i$ with the semi-implicit Euler method:
\begin{equation}
\begin{aligned}
\vec{v}_i &\gets \vec{v}_i + \Delta t\,\vec{f}_i / m\\
\vec{r}_i &\gets \vec{r}_i + \Delta t\,\vec{v}_i
\end{aligned}
\end{equation}
where $\Delta t$ is the timestep size. Since we eliminated all masses from the program code ($m=1$), the vertex forces $\vec{f}_i$ are effectively accelerations.

\begin{table*}
\centering
\caption{\textbf{Model parameters.} In the parameter dimension, M represents mass, L length, T time. Masses are normalized in our implementation ($\text{M}=1$). Default values produce an exponentially growing epithelial tissue.}
\label{tab:2}
\begin{tabular}{lrccl}
\toprule
Symbol & Default value & Constraints & Dimension & Description\\
\midrule
\multicolumn{5}{c}{\textbf{Geometric parameters}}\\
\midrule
$h$ & 0.01 & $>0$ & L & Edge thickness\\
$l_\mathrm{min}$ & $0.02$ & $\geq0$ & L & Minimum edge length\\
$l_\mathrm{max}$ & $0.2$ & $>2l_\mathrm{min}$ & L & Maximum edge length\\
$Q$ & $1$ & $>0$ & --- & Target isoparametric ratio\\
$\alpha$ & 1 & --- & L\textsuperscript{2}/T & Area growth rate\\
$A_\mathrm{min}$ & $0$ & $\geq0$ & L\textsuperscript{2} & Minimum polygon area (for cell removal)\\
$A_\mathrm{max}$ & $\pi$ & $>A_\mathrm{min}$ & L\textsuperscript{2} & Maximum polygon area (for cell division)\\
$s_\mathrm{h}$ & $0.01$ & $\geq0$ & L & Adhesion hardening zone size\\
$s_\mathrm{s}$ & $0.01$ & $\geq0$ & L & Adhesion softening zone size\\
$\theta$ & $0$ & $0\leq\theta\leq 1$ & --- & Fusion threshold\\
\midrule
\multicolumn{5}{c}{\textbf{Material parameters}}\\
\midrule
$k_\mathrm{a}$ & $10^5$ & $\geq0$ & M/L\textsuperscript{2}T\textsuperscript{2} & Area stiffness\\
$\gamma$ & $10^3$ & --- & ML/T\textsuperscript{2} & Boundary line tension\\
$k_\mathrm{l}$ & $10^4$ & $\geq0$ & ML/T\textsuperscript{2} & Tensile rigidity\\
$k_\mathrm{b}$ & 0 & $\geq0$ & ML\textsuperscript{3}/T\textsuperscript{2} & Bending rigidity\\
$k_\mathrm{r}$ & $10^7$ & $\geq0$ & 1/MT\textsuperscript{2} & Repulsion stiffness\\
$k_\mathrm{h}$ & $10^6$ & $\geq0$ & 1/MT\textsuperscript{2} & Adhesion stiffness\\
$\mu$ & 0 & $\geq0$ & --- & Dynamic Coulomb friction coefficient\\
$\rho$ & 0 & --- & M/L\textsuperscript{2} & Fluid mass density\\
\midrule
\multicolumn{5}{c}{\textbf{Other parameters}}\\
\midrule
$g$ & 0 & --- & L/T\textsuperscript{2} & Gravitational acceleration\\
$g_\mathrm{l}$ & 0 & --- & L/T\textsuperscript{2} & Edge gravitational acceleration\\
$c_\mathrm{v}$ & $10$ & $\geq0$ & M/T & Viscous damping coefficient\\
$c_\mathrm{c}$ & $30$ & $\geq0$ & 1/T & Contact damping coefficient\\
$c_\mathrm{d}$ & 0 & $\geq0$ & --- & Drag coefficient\\
$\Delta t$ & $10^{-4}$ & $>0$ & T & Time increment\\
$m$ & 1 & fixed & M & Vertex mass (not in the code)\\
\bottomrule
\end{tabular}
\end{table*}

\section{Applications}

\program{} can be applied to a broad range of 2D elastic or free surface problems governed by effectively 23 physical parameters---ten geometrical, eight material, and five others (not counting the numerical time increment). Table~\ref{tab:2} lists them all, including constraints that need to be respected, and a set of simple default values that quickly produce usable simulation output and may serve as a starting point to find working parameters for specific applications. We now showcase a selection of scenarios that can be represented with these 23 model parameters.

\subsection{Biological tissues}

We start with a series of biological examples. The classical application is a growing epithelial monolayer tissue. Fig.~\ref{fig:2}A and Movie 1 show a simulation with about 1000 cells with thin membranes ($h=0.01$) whose equilibrium shape is a circle ($Q=1$). The cells are adhesive, mildly compressible, and their membranes are governed by cortical tension and relatively weak line elasticity (Table~\ref{tab:3}). Bending, fusion, friction, gravity, and drag are disabled by setting their respective coefficients to zero. The simulation starts with a single cell with unit radius. Cells divide upon reaching an area of $A_\mathrm{max}=\pi$, and grow at a rate drawn from a normal distribution for each cell independently. For demonstration purposes, negative growth rates are allowed here, and cells undergo apoptosis (i.e., they are removed) when they become small. As shown in the closeup in Fig.~\ref{fig:2}A, the cells end up being non-convex, with curved membranes. Simulations like these are widespread in computational biology and in essence reproducible with existing software \cite{Rejniak:2005, Dillon:2008, Jamali:2010, Merks:2011, Mkrtchyan:2014, Tanaka:2015, Cooper:2017, Conradin:2021, Brown:2021, Tervonen:2023}; with \program{}, they now run in a few minutes.

\program{} is also suited to study active motion (motility), migration, collective behavior, or aggregation/segregation. Although this is not implemented in the supplied default code, a localized attractor, for example, can be added with a single line of additional code, to model chemotaxis, durotaxis, or similar phenomena. In Fig.~\ref{fig:2}B and Movie 2, we picked 20 cells at random (red) and let them migrate toward a single point with a radial attractive potential. With such simulations, open 2D problems in cell motility \cite{Ziebert:2016} may be addressable efficiently.

Also non-confluent, structured tissues can be represented. Fig.~\ref{fig:2}C shows an example simulation with epithelial vesicles (brown cells) arranged in monolayer rings surrounding a luminar region (light blue). The parametric setup is similar to the previous examples (Table~\ref{tab:3}), with the notable exception that we set the isoparametric ratio to that of squares, $Q=4/\pi$, such that in absence of bending and cortical tension, the cells are mechanically satisfied with a rectangular shape.

\begin{table*}
\centering
\caption{\textbf{Model parameters for the simulations shown in Figs.~\ref{fig:2}--\ref{fig:4}.} See Table \ref{tab:2} for an explanation of the parameters.}
\label{tab:3}
\setlength{\tabcolsep}{0.28em}
\begin{tabular}{@{}lrrrrrrrrrrrrrrrrrrrrrrrr@{}}
\toprule
Fig. & $h$ & $l_\mathrm{min}$ & $l_\mathrm{max}$ & $Q$ & $\alpha$ & $A_\mathrm{min}$ & $A_\mathrm{max}$ & $k_\mathrm{a}$ & $\gamma$ & $k_\mathrm{l}$ & $k_\mathrm{b}$ & $k_\mathrm{r}$ & $k_\mathrm{h}$ & $s_\mathrm{h}$ & $s_\mathrm{s}$ & $\theta$ & $\mu$ & $g$ & $g_\mathrm{l}$ & $\rho$ & $c_\mathrm{v}$ & $c_\mathrm{c}$ & $c_\mathrm{d}$ & $\Delta t$\\
\midrule
2A & $0.01$ & $0.02$ & $0.2$ & $1$ & var\textsuperscript{\textdagger} & $0.1$ & $\pi$ & $10^5$ & $10^4$ & $10^4$ & $0$ & $10^7$ & $10^6$ & $0.01$ & $0.01$ & $0$ & $0$ & $0$ & $0$ & $0$ & $20$ & $30$ & $0$ & $10^{-4}$\\
2B & $0.01$ & $0.02$ & $0.1$ & $1$ & $0$ & $0$ & $\infty$ & $10^5$ & $10^4$ & $10^5$ & $0$ & $10^7$ & $0$ & $0$ & $0$ & $0$ & $0$ & $0$ & $0$ & $0$ & $20$ & $30$ & $0$ & $10^{-4}$\\
2C & $0.01$ & $0.02$ & $0.2$ & $4/\pi$ & --- & $0$ & $\infty$ & $10^5$ & $10^5$ & $10^4$ & $0$ & $10^7$ & $10^6$ & $0.01$ & $0.01$ & $0$ & $0$ & $0$ & $0$ & $0$ & $5$ & $30$ & $0$ & $10^{-4}$\\
2D & $0.02$ & $0.02$ & $0.04$ & $3$ & --- & $0$ & $\infty$ & $10^5$ & $0$ & $10^5$ & $1$ & $10^7$ & $1$ & $0.2$ & $0.2$ & $0$ & $0$ & $0$ & $0$ & $0$ & $10$ & $30$ & $0$ & $10^{-4}$\\
2E & $0.05$ & $0.01$ & $0.04$ & $1$ & $0$ & $0$ & $\infty$ & $0$ & $0$ & $10^5$ & $10^2$ & $10^7$ & $0$ & $0$ & $0$ & $0$ & $0$ & $0$ & $10$ & $0$ & $2$ & $100$ & $0$ & $10^{-4}$\\
2F & $0.01$ & $0.01$ & $0.04$ & --- & $0$ & $0$ & $\infty$ & $10^7$ & $10^3$ & $0$ & $0$ & $10^7$ & $25$ & $0.02$ & $0.02$ & $0$ & $0$ & $0$ & $2$ & $0$ & $0.1$ & $30$ & $0$ & $10^{-4}$\\
2G & $0.01$ & $0.01$ & $0.03$ & --- & $0$ & $0$ & $\infty$ & $10^7$ & $10^2$ & $0$ & $0$ & $10^5$ & $0$ & $0$ & $0$ & $0.85$ & $0$ & $1$ & $0$ & $10^3$ & $1$ & $30$ & $1$ & $10^{-4}$\\
2H & $0.04$ & $0.05$ & $0.2$ & $1$ & $0$ & $0$ & $\infty$ & $10^5$ & $10^4$ & $10^4$ & $0$ & $5\!\times\!10^7$ & $10^4$ & $0.02$ & $0.02$ & $0$ & $0.5$ & $0$ & $10$ & $0$ & $1$ & $30$ & $0$ & $10^{-4}$\\
\midrule
3 & $0.01$ & $0.02$ & $0.1$ & $1$ & $1$ & $0$ & var\textsuperscript{\textdaggerdbl} & $10^5$ & $10^5$ & $10^4$ & $0$ & $10^7$ & $10^6$ & $0.01$ & $0.01$ & $0$ & $0$ & $0$ & $0$ & $0$ & $2$ & $30$ & $0$ & $10^{-4}$\\
\midrule
4A,B & $0.01$ & $0.02$ & $0.2$ & $1$ & $1$ & $0$ & $\pi$ & $10^4$ & $10^5$ & $10^4$ & $0$ & $10^7$ & $10^6$ & $0.01$ & $0.01$ & $0$ & $0$ & $0$ & $0$ & $0$ & $2$ & $30$ & $0$ & $10^{-4}$\\
4C,D & $0.01$ & $0.02$ & $0.2$ & $24/\sqrt3$ & $0$ & $0$ & $\infty$ & $10^4$ & $10^5$ & $10^4$ & $0$ & $10^7$ & $10^6$ & $0.01$ & $0.01$ & $0$ & $0$ & $0$ & $0$ & $0$ & $2$ & $30$ & $0$ & $10^{-4}$\\
\midrule[\heavyrulewidth]
\multicolumn{25}{@{}l@{}}{\textsuperscript{\textdagger}Randomly drawn from a normal distribution with a mean and standard deviation of $1$ (including negative values).}\\
\multicolumn{25}{@{}l@{}}{\textsuperscript{\textdaggerdbl}Randomly drawn from a lognormal distribution with mean $\pi$ and coefficient of variation $0.3$.}\\
\end{tabular}
\end{table*}

\subsection{Amorphous materials}

\begin{figure*}
	\centering
	\includegraphics[width=\linewidth]{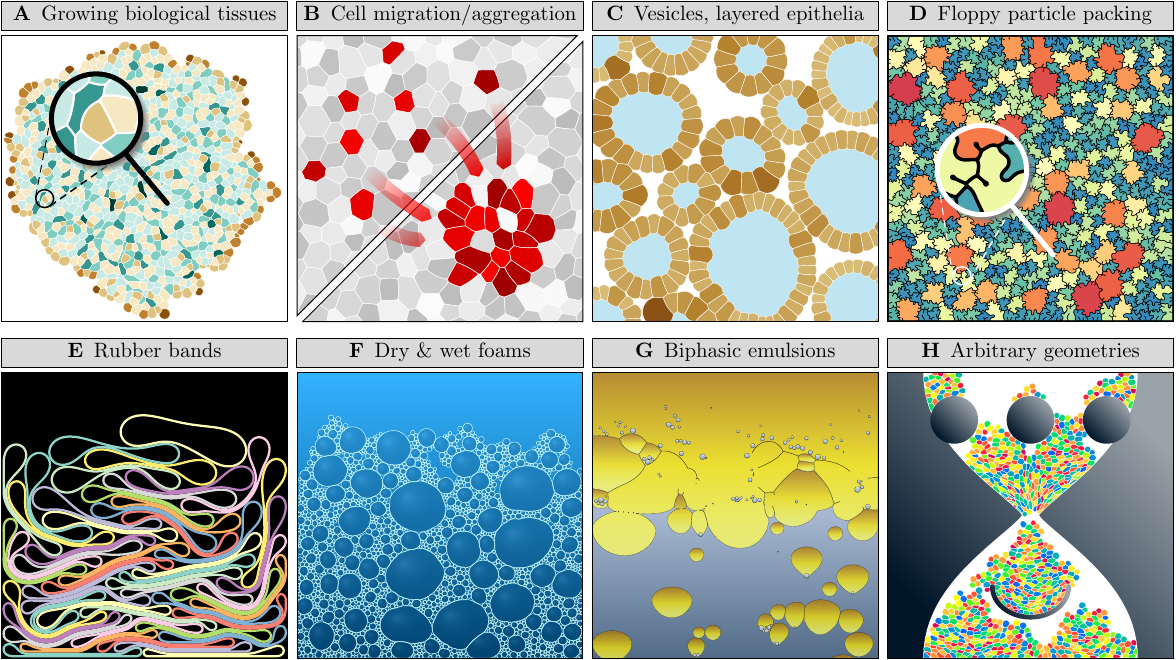}
	\caption{\textbf{Scope of representable systems.}
    Example simulations ranging from the growth of biological monolayer tissues (A), active matter such as migrating and aggregating cells (B), vesicles and layered epithelia (C), highly amorphous materials composed of floppy particles with excess perimeter (D), packing of elastic rubber bands (E), foams with various degrees of fluid content and bubble dispersity (F), emulsions with arbitrarily nested, fusing and separating drops (G), to arbitrarily complex and curved domain shapes (H). Model parameters are specified in Table~\ref{tab:3}.
	}
	\label{fig:2}
\end{figure*}

A further field of application is the study of amorphous materials and their dense packing, jamming, rheology, etc., in the spirit of recent numerical work \cite{Boromand:2018,Boromand:2019,Treado:2021}. To illustrate an extreme case, we close-packed an ensemble of polydisperse floppy particles with excess circumference ($Q=3$) (Fig.~\ref{fig:2}D). For this simulation, the particles were set to be mildly compressible, weakly adhesive, with a slightly thicker boundary ($h=0.02$) driven by line elasticity and weak bending rigidity (Table~\ref{tab:3}). Line tension, gravity, friction etc. were disabled. The resulting configuration exhibits interlocked particles with tight local folds to accommodate the long boundaries. Although their genesis may be different (i.e., without internal joints), these particle shapes resemble those of the the puzzle-shaped epidermal cells of \textit{Arabidopsis thaliana} leaves \cite{Sapala:2018}.

\subsection{Elastic bands}

To showcase an application dominated by bending and gravitational forces, we simulated the downfall and packing of 100 thick elastic bands. Boundary gravity ($\rho_\mathrm{l}=10$) compresses the initially vertically piled-up circular rings to the point where they partially align into parallel bundles (Fig.~\ref{fig:2}E). For an animated version, see Movie 3. For this simulation, we set the area stiffness, line tension, and adhesion parameter values to zero, but increased the thickness and bending rigidity (Table~\ref{tab:3}). \program{} thus enables simulations of the packing of elastic rods (previously performed on linear rods \cite{Stoop:2008, Vetter:2013}) with circular hoops.

\begin{figure*}
	\centering
	\includegraphics[width=\linewidth]{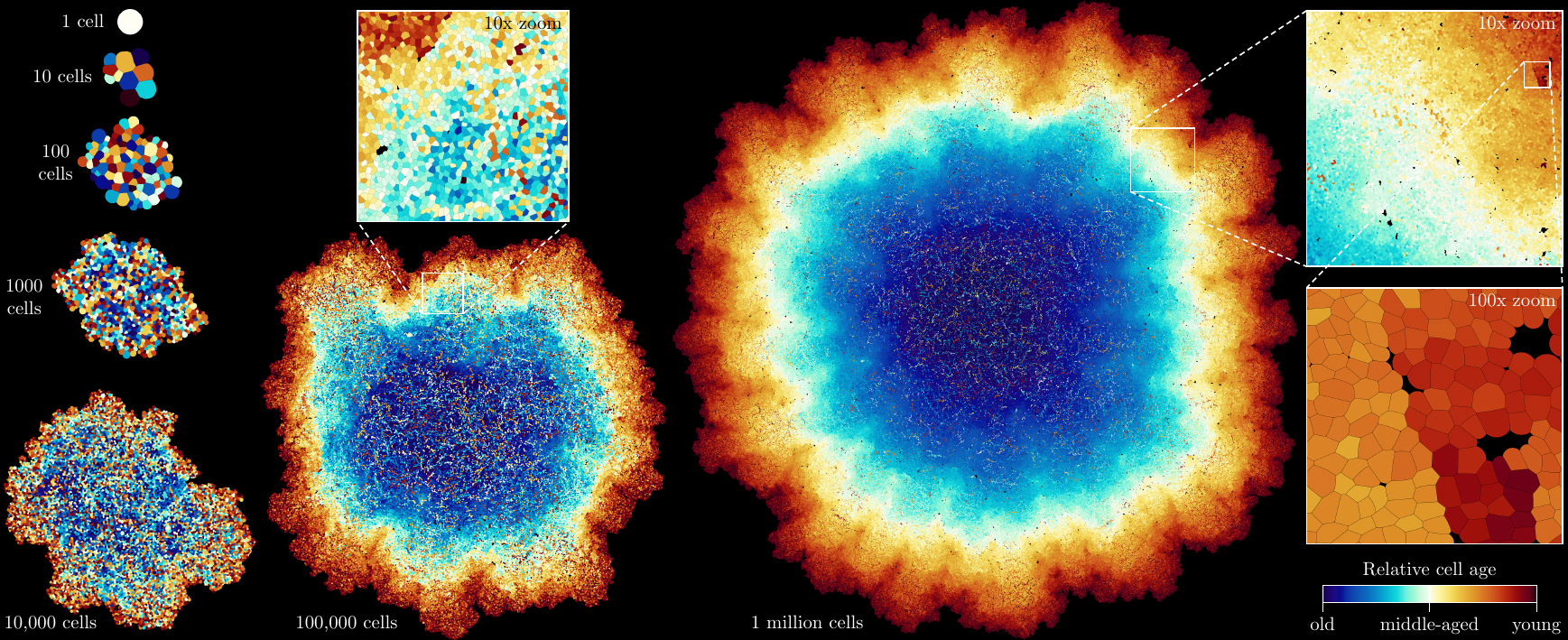}
	\caption{\textbf{Organ-scale biological tissue simulation.} The color represents the relative cell age in the entire tissue; upon division, one daughter cell starts at age 0, while the other inherits its mother's age. The snapshots are not to scale; each uses its own color scale. Model parameters are specified in Table~\ref{tab:3}. 
	}
	\label{fig:3}
\end{figure*}

\subsection{Foams and emulsions}

Also belonging to the class or amorphous solids, but worth mentioning separately here for their historical significance (cf.\ Table \ref{tab:1}), are foams and emulsions. A relatively wet foam is shown in Fig.~\ref{fig:2}F, produced by letting a polydisperse collection of nearly incompressible bubbles, initially circular in shape and randomly placed in the plane, drop under the action of gravity. Governed by surface tension and weak cohesion (Table~\ref{tab:3}), the bubbles deform and coalesce into a foamy structure with a free upper surface (Movie 4).

Perhaps technically the most challenging application in this exhibition is the simulation of a binary emulsion in which oil drops in an aqueous medium freely fuse and split (Fig.~\ref{fig:2}G). In this simulation, we use a mass density difference of $\rho=10^3$ between the two phases to let the oil drops float upward in response to hydrostatic pressure, slowed down by drag ($c_\mathrm{d}=1$) (Table~\ref{tab:3}). With a fusion threshold of $\theta=0.85$, oil drops pressing against each other merge, broken bubble boundaries retract driven by interfacial tension, and a free water-oil interface forms naturally, separating the two phases (Movie 5). Note that \program{} poses no limit to the recursion depth in the drop-in-drop cascade. For demonstration purposes, we placed water droplets in the oil drops here, which are then joined by further droplets that spontaneously form from the interfaces breaking up when drops fuse. The entire simulation completes in about 10 minutes. Simulations of this kind may for instance be used to advance earlier 2D foaming studies \cite{Korner:2002, Ataei:2021} to a regime involving larger topological rearrangements.

\subsection{Complex geometries}
\label{sec:geometry}

\program{} offers native support for arbitrarily shaped geometrical domains and obstacles. We demonstrate this with a simulation of about 1000 soft sticky particles dropping in an hourglass-like geometry with four additional curved objects in the way (Fig.~\ref{fig:2}H, Movie 6). For this simulation, we also turned on friction ($\mu=0.5$, Table~\ref{tab:3}). Rigid objects, which can serve both as confining bounds and obstacles, are implemented in \program{} with a simple, flexible approach requiring minimal coding: They are treated as normal particles, with two exceptions: They are not coarsened to maintain a preset resolution of curved regions (but refined to ensure $l_i\leq l_\mathrm{max}$, enabling efficient local contact detection), and their vertex positions are not updated to make them immobile. This delegates the geometric definition of rigid domain boundaries to the input file (see Sec.~\ref{sec:usage}), and renders specialized collision handling with primitive shapes unnecessary.

\subsection{Large-scale simulation}

Finally, we return to a biophysical application to demonstrate the scale \program{} can reach. A strongly proliferative, cancer-like tissue is grown from a single cell to a million cells (Fig.~\ref{fig:3}). To our knowledge, this is the first published report of a comparable simulation exceeding \num{10000} cells with high boundary resolution. Cell boundaries have a mean resolution of $M=(1/N)\sum_{p=1}^NM_p\approx72$ vertices, totaling in about 72 million simulated vertices in the final tissue. To avoid excessive pressure buildup in the interior of the tissue due to the fast growth speed, cells are grown only if $A_p/A_{p,0}>90\%$ in this simulation. At around $N\approx10^4$ cells (Fig.~\ref{fig:3}, bottom left), we observe a transition from spatially uniform cell proliferation to a radial gradient in the relative cell age, with a central region where the compressed cells (blue) are able to grow to their division area only sporadically, and a rugged rim consisting mainly of very young, proliferative cells (red). Closeups reveal regions exhibiting sharp cell age boundaries within the tissue. This example demonstrates how organ-scale emergent features can be studied with \program{} without relinquishing cellular fidelity. See Movie 7 for an animated version.

\section{Parallelization and computational efficiency}

Up to caching effects, the total serial time complexity for simulations with \program{} is $\mathcal{O}(NMD)$, where $N$ is the number of polygons, $M$ is the mean number of vertices per polygon, and $D=N_\mathrm{f}N_\mathrm{s}$ is the simulated time duration in units of steps.  The number of output frames, $N_\mathrm{f}$, and the number of time steps per frame, $N_\mathrm{s}$, are simulation output parameters that can be set in the code. The runtime of the program thus scales linearly with the total number of vertices, $NM$. Higher spatial resolution, temporal resolution, or number of simulated polygons affect the runtime proportionally.

\program{} is parallelized for shared-memory computers, using OpenMP. With just six \texttt{omp parallel for} directives, the main loops for bounding box computation, force computation, polygon interactions, fusion testing, remodeling, and time propagation are parallelized without explicit use of threading expressions, leaving the serial flow of the source code untouched. Three relevant code segments are not parallelized: I/O, the construction of neighbor lists (because of speedup limitations \cite{Halver:2016}), and the execution of topological changes. The latter is due to changes in polygon count and the involvement of polygons in multiple topological changes in a single timestep, which makes it challenging to apply simple parallelization strategies.

\begin{figure}
	\centering
	\includegraphics[width=\linewidth]{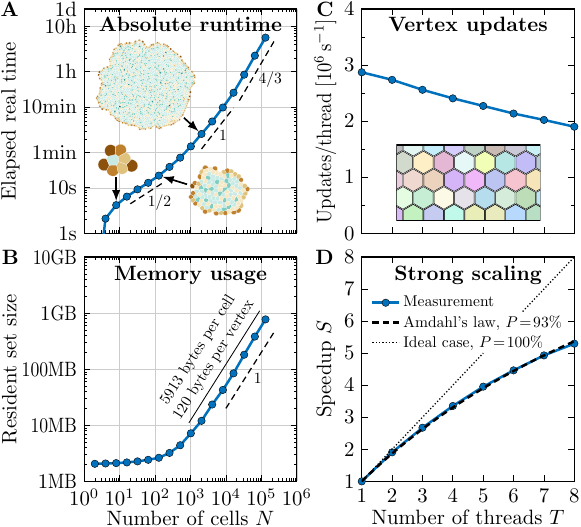}
	\caption{\textbf{Computational performance.}
	\textbf{A, B} Runtime and memory usage for a biological tissue simulation (depicted, cells colored by neighbor number) with exponential growth starting from a single cell. Number of threads $T=8$, mean number of vertices per cell $M\approx50$. Asymptotic time and memory complexity exponents are indicated (dashed lines).
	\textbf{C, D} Parallel scaling in a hexagonal dense packing simulation without growth or cell division (inset). $N=10^5$ cells with $M=48$ vertices each. Amdahl's law: $S=1/(1-P+P/T)$, where $P$ is the parallel fraction.
	Runtime and scaling tests were performed on a MacBook Pro with an Intel Core i9-9880H CPU (8 cores, 2.30 GHz). I/O was excluded from the measurement. Model parameters are specified in Table~\ref{tab:3}.
	}
	\label{fig:4}
\end{figure}

To assess the computational efficiency in absolute terms, we measured the wallclock time in a typical biological tissue simulation with exponential growth on an ordinary modern laptop computer using 8 threads. Starting from a single proliferative cell, a tissue was produced by successive cell growth and division, similar to Figs.~\ref{fig:2}A and \ref{fig:3}, and the elapsed time was recorded for each doubling of the cell count (Fig.~\ref{fig:4}A). For a few hundred cells, we observe square-root time complexity in the number of grown cells $N$, presumably due to cache locality. In an intermediate range with some thousand cells, the runtime increases linearly with the number of cells, until about $N=10^4$, which is reached in about 12 minutes at a resolution of $M\approx50$ vertices per cell. Beyond that, the runtime is superlinear with power-law exponent $4/3$ due to a slow-down of proliferation caused by pressure building up in the interior of the tissue, which can in principle be avoided with lower growth rates. A time complexity of $\mathcal{O}(N^{4/3})$ is also observed in similar simulations in 3D \cite{Runser:2023}. A tissue consisting of $N=2^{17}=\num{131074}$ cells is grown in less than six hours. For this performance benchmark, we excluded I/O and cell fusion testing.

In an analogous simulation, we measured the memory usage under Red Hat Enterprise 7.9 and found the expected asymptotically linear scaling (Fig.~\ref{fig:4}B). The total memory complexity is thus $\mathcal{O}(NM)$. Each vertex uses 72 bytes of memory in IEEE double precision on a 64bit architecture: 7 floating-point numbers for the vertex position, velocity, acceleration, and target length of the adjacent edge, and 2 indices or pointers for the polygon affiliation and forward vertex linkage in the space partitioning grid. Including per-polygon data of 64 bytes each, the spatial grid, temporaries, and bookkeeping overhead, we measured an overall memory requirement of about 120 bytes per vertex in a typical tissue growth scenario. A simulation with $N=10^4$ polygons, consisting of $M=50$ vertices each, therefore requires approximately 60 MB of main memory in total. A large high-fidelity simulation with $N=10^6$ polygons, consisting of $M=100$ vertices each, will consume about 12 GB, and will thus still easily fit into the main memory of most ordinary modern computers.

\program{} scales well on multi-core CPUs. The number of vertex updates that are performed per second and per thread in a large hexagonal close-packing setting (Fig.~\ref{fig:4}C, inset), in which all interior edges are in active contact with neighboring polygons, drops by only a third from almost 3 million in serial mode to about 2 million with 8 threads (Fig.~\ref{fig:4}C). We observe the strong-scaling behavior to follow Amdahl's law very closely, with a speedup of $S=1/(1-P+P/T)$, where $P=93\%$ is the parallel portion and $T$ the number of threads (Fig.~\ref{fig:4}D). The remaining 7\% are largely due to the serial placement of vertices into the spatial partitioning grid.

\section{Usage instructions}
\label{sec:usage}

\program{} was developed to be highly portable, but it requires a compiler compatible with the C++11 standard. For multithreading support, the OpenMP 3.1 specification or newer is additionally required. We tested it using GCC v4.8.2--12.2.0, Clang/LLVM v3.6.0--15.0.7, and ICC v14.0.1--19.1.0. Compiling \program{} is as simple as running
\begin{verbatim}
g++ -fopenmp -O3 -o polyhoop polyhoop.cpp
\end{verbatim}
or an equivalent command. \program{} can be compiled for serial execution by omitting the OpenMP compile option.

To run a simulation, set the desired parameters in \texttt{polyhoop.cpp} (lines 13--42), compile it, and execute the binary by typing
\begin{verbatim}
OMP_NUM_THREADS=8 ./polyhoop
\end{verbatim}
or similar. If \program{} is compiled with OpenMP support and the number of threads is not specified, the selection of a suitable number of threads is delegated to OpenMP.

\program{} reads in a mandatory input file named \texttt{ensemble.off} from the current working directory, specifying the initial configuration in GeomView's Object File Format (OFF) \cite{OFF}. Since OFF specifies vertex coordinates in three dimensions $x$,$y$,$z$, but \program{} uses only two ($x$ and $y$), the $z$ coordinate is used to indicate the phase of the polygon containing the vertex. $z$ is expected to be an integer that, if odd, denotes phase $\phi_p=1$, and if even, $\phi_p=-1$. The input file only specifies vertex positions; velocities are initialized to zero. The first $N_\mathrm{r}$ polygons in the input file are treated as rigid obstacles (see Sec.~\ref{sec:geometry}). $N_\mathrm{r}$ can be set in the source file.

The creation of such input files representing suitable initial starting conditions for different applications is left to the user. The supplied default input file \texttt{ensemble.off} contains a single 40-node circular hoop with unit radius, as used in the tissue growth simulations showcased in Figs.~\ref{fig:2}A, \ref{fig:3} and \ref{fig:4}A,B. The default parameter setup (Table~\ref{tab:2}) grows a 128-cell tissue out of this single initial cell in about 20 seconds on an Intel Core i9-9880H CPU using four threads.

\program{} writes the simulated configurations at specified time intervals (every $N_\mathrm{s}$ timesteps) to the current working directory, naming them \texttt{frame\%06d.vtp}. With each output frame, a single-line status report is printed to the command line. The output files are in the VTK ASCII polygon format \cite{VTK}. For each polygon in the ensemble, they store the area, perimeter and coordination number (the number of other polygons each polygon is separated from by a distance shorter than $s_\mathrm{h}+s_\mathrm{s}$). The simulation output can be visualized and animated by opening the VTK file series in ParaView (\url{https://paraview.org}). For non-convex polygons, the \texttt{Extract Surface} and \texttt{Triangulate} filters need to be applied to ensure correct visual representation. Similar to the input file, the phase $\phi_p$ of each polygon is stored in the unused $z$ coordinate of the output VTK files, as a binary flag $z=(\phi_p+1)/2$.

\section{Discussion and outlook}

\program{} is an exceptionally lightweight standalone program that enables the simulation of a wide variety of dynamic phenomena governed by constrained visco-elastic hoops in 2D, such as monolayer tissues, emulsions, or other ensembles of soft deformable particles. \program{} stands out of the crowd of related model implementations \cite{Tanaka:2015,Boromand:2018,Conradin:2021} in mainly three aspects:
\begin{itemize}
\item \textit{Topological flexibility.} Hoops/polygons can be broken up and rejoined, tissues can grow together or detach, new cells can emerge through cell division or disappear by death or extrusion, bubbles can merge or split up. 
\item \textit{Computational performance.} \program{} is designed to be particularly lean and efficient, making it the first of its kind (to our knowledge) to enable large-scale simulations with hundreds of thousands or even millions of polygons with high spatial resolution. Biological tissue simulations of the size of entire organs can be simulated on customary computers, without requiring access to the brute force of cluster computing infrastructure or GPUs.
\item \textit{Accessibility.} \program{} is completely free and open-source. Comprising only just above 700 lines of compact, simple, commented C++ code, it is easy to handle, extend and embed in other programs. Importantly, it has no dependencies, making it exceptionally independent and portable. Thanks to the 3-clause BSD license under which \program{} is published, both commercial and non-commercial uses are permitted.
\end{itemize}

With \program{}, computer simulations of 2D tissue morpogenesis enter a regime in which open questions at macroscopic length scales can be addressed numerically with high spatial resolution of individual cells. Complex intertwined cell shapes have recently been discovered in pseudostratified epithelia \cite{Gomez:2021,Iber:2022a}, and their role in organogenesis and patterning is a subject of ongoing research \cite{Iber:2022b}. Interkinetic nuclear migration, a process in which migrating nuclei can strongly deform the cells, remains largely mysterious \cite{Spear:2012}. Studying these phenomena in silico requires models that offer a degree of geometrical flexibility that allows to link cell shape to tissue function and morphology. \program{} is a high-throughput tool that enables such simulations not only with some dozens of cells, but with hundreds of thousands.

A particular strength of \program{}, which also happens to be a core application it was developed for, is the growth of proliferative tissue to developmentally or medically relevant sizes, and other morphogenetic events on that scale. The cell count of the \textit{Drosophila} wing disc (\num{50000} \cite{Buchmann:2014}), for example, one of the most-studied organs in developmental biology, is reached in about 1.5 hours on a typical modern 8-core CPU, with each cell discretized by 50 vertices, starting from a single cell (Fig.~\ref{fig:4}A). Another natural application is macroscopic growth of malignant tumors (Fig.~\ref{fig:3}), which are linked to aberrant cell shape and impairment of structural tissue integrity \cite{Wu:2020}.

With polygon fusion, \program{} adds a feature to the pool of 2D free-boundary evolvers that enables the simulation of topological transitions occurring in a variety of bubbly systems, from emulsions to biological tissues. Cell fusion is crucial in tissue development \cite{Hernandez:2017}, but not commonly included in cell-based simulations of tissue development. Our program is well suited to simulate such processes. The current implementation uses a simple geometrical indentation criterion with parameter $\theta$ to trigger fusion, which is equivalent to a compressive stress threshold. With minimal changes to the source code, cells could be made to fuse based on other conditions such as time, position, cell size and shape, cell type, and potentially biochemical signals.

Several further extensions and generalizations of the present model are easy to recognize. For simplicity, all polygons in \program{} share the same constitutive relationships and material parameters (apart from the area growth rate and division area threshold, which are drawn from random distributions in the supplied code). It is straightforward to make these properties members of the individual polygons, or even vertices or edges if needed, with only a few lines of code to be changed. This will allow to simulate different types of particles in the ensemble, or non-uniform or inhomogeneous material behavior such as differential cell adhesion and cell polarity. Other conceivable extensions include active or Brownian motion, tension fluctuations \cite{Kim:2021,Krajnc:2021}, and the coupling to fluid dynamics or reaction-diffusion solvers, as offered by other frameworks for applications in biology \cite{Tanaka:2015,Conradin:2021}, among others. For the simulation of realistic emulsions with deeply nested phases, it may also be desirable to make the hydrostatic pressure dependent on the vertical immersion depth in the surrounding phase rather than on the global position, as currently implemented. For applications in which the hoops represent actual physical entities (such as elastic threads or cell membranes) rather than just marking the boundaries between immiscible fluid phases, it could be useful to extend \program{} to allow for the rupture of hoops without immediate resealing. That is to say, allowing the polygons to have holes; to be open lines with endpoints rather than closed hoops.

On the numerical side, certain room may exists for further performance improvements. The computational performance of the present implementation is largely memory access-limited, implying that it could potentially benefit from parallelization for distributed memory systems, for instance with MPI. Moreover, a more elaborate bookkeeping of nearby vertex pairs, for example with Verlet lists \cite{Verlet:1967}, could speed up simulations further, considering that about 60--90\% of the runtime is spent in contact detection in typical tissue growth scenarios. For \program{}, we have deliberately resorted to a simple and compact solution using repeatedly computed linked cell lists and parallelization with OpenMP precompiler directives.

Finally, we wish to make the interested reader aware of model developments that strive to offer functionality similar to that of \program{} in 3D \cite{Da:2014,Madhikar:2018,Liedekerke:2020,Wang:2021,Torres:2022,Runser:2023,Okuda:2023}, albeit mostly without some of the topological transitions implemented here, and naturally at substantially greater computational cost and geometrical complexity. 

\section*{Acknowledgement}

We thank Marius Almanst\"{o}tter for help with parameter testing, as well as Marco Meer and the group of Bastien Chopard for valuable discussions. Financial support from the Swiss National Science Foundation by Sinergia grant no.\ 170930 is gratefully acknowledged.

\section*{Competing Interests}

The authors declare that they have no competing interests.


\begin{thebibliography}{81}
\providecommand{\natexlab}[1]{#1}
\providecommand{\url}[1]{\texttt{#1}}
\expandafter\ifx\csname urlstyle\endcsname\relax
  \providecommand{\doi}[1]{doi: #1}\else
  \providecommand{\doi}{doi: \begingroup \urlstyle{rm}\Url}\fi

\bibitem[Kawasaki et~al.(1989)Kawasaki, Nagai, and Nakashima]{Kawasaki:1989}
K.~Kawasaki, T.~Nagai, and K.~Nakashima.
\newblock Vertex models for two-dimensional grain growth.
\newblock \emph{Phil. Mag. B}, 60:\penalty0 399--421, 1989.
\newblock \doi{10.1080/13642818908205916}.

\bibitem[Weliky and Oster(1990)]{Weliky:1990}
M.~Weliky and G.~Oster.
\newblock {The mechanical basis of cell rearrangement I. Epithelial
  morphogenesis during Fundulus epiboly}.
\newblock \emph{Development}, 109:\penalty0 373--386, 1990.
\newblock \doi{10.1242/dev.109.2.373}.

\bibitem[Graner and Sawada(1993)]{Graner:1993}
F.~Graner and Y.~Sawada.
\newblock {Can Surface Adhesion Drive Cell Rearrangement? Part II: A
  Geometrical Model}.
\newblock \emph{J. Theor. Biol.}, 164:\penalty0 477--506, 1993.
\newblock \doi{10.1006/jtbi.1993.1168}.

\bibitem[Nagai and Honda(2001)]{Nagai:2001}
T.~Nagai and H.~Honda.
\newblock A dynamic cell model for the formation of epithelial tissues.
\newblock \emph{Philos. Mag. B}, 81:\penalty0 699--719, 2001.
\newblock \doi{10.1080/13642810108205772}.

\bibitem[Farhadifar et~al.(2007)Farhadifar, R\"{o}per, Aigouy, Eaton, and
  J\"{u}licher]{Farhadifar:2007}
R.~Farhadifar, J.-C. R\"{o}per, B.~Aigouy, S.~Eaton, and F.~J\"{u}licher.
\newblock {The Influence of Cell Mechanics, Cell-Cell Interactions, and
  Proliferation on Epithelial Packing}.
\newblock \emph{Curr. Biol.}, 17:\penalty0 2095--2104, 2007.
\newblock \doi{10.1016/j.cub.2007.11.049}.

\bibitem[Hufnagel et~al.(2007)Hufnagel, Teleman, Rouault, Cohen, and
  Shraiman]{Hufnagel:2007}
L.~Hufnagel, A.~A. Teleman, H.~Rouault, S.~M. Cohen, and B.~I. Shraiman.
\newblock On the mechanism of wing size determination in fly development.
\newblock \emph{Proc. Natl. Acad. Sci. U.S.A.}, 104:\penalty0 3835--3840, 2007.
\newblock \doi{10.1073/pnas.0607134104}.

\bibitem[Vetter et~al.(2019)Vetter, Kokic, G\'{o}mez, Hodel, Gjeta, Iannini,
  Villa-Fombuena, Casares, and Iber]{Vetter:2019}
R.~Vetter, M.~Kokic, H.~F. G\'{o}mez, L.~Hodel, B.~Gjeta, A.~Iannini,
  G.~Villa-Fombuena, F.~Casares, and D.~Iber.
\newblock Aboav-weaire's law in epithelia results from an angle constraint in
  contiguous polygonal lattices.
\newblock \emph{BioRxiv}, 2019.
\newblock \doi{10.1101/591461}.

\bibitem[Bi et~al.(2015)Bi, Lopez, Schwarz, and Manning]{Bi:2015}
D.~Bi, J.~H. Lopez, J.~M. Schwarz, and M.~L. Manning.
\newblock {A density-independent rigidity transition in biological tissues}.
\newblock \emph{Nat. Phys.}, 11:\penalty0 1074--1079, 2015.
\newblock \doi{10.1038/nphys3471}.

\bibitem[Kim et~al.(2021)Kim, Pochitaloff, Stooke-Vaughan, and
  Camp\`{a}s]{Kim:2021}
S.~Kim, M.~Pochitaloff, G.~Stooke-Vaughan, and O.~Camp\`{a}s.
\newblock {Embryonic Tissues as Active Foams}.
\newblock \emph{Nat. Phys.}, 2021.
\newblock \doi{10.1038/s41567-021-01215-1}.

\bibitem[Weaire(1992)]{Weaire:1992}
D.~Weaire.
\newblock Some lessons from soap froth for the physics of soft condensed
  matter.
\newblock \emph{Phys. Scr.}, 1992:\penalty0 29--33, 1992.
\newblock \doi{10.1088/0031-8949/1992/T45/006}.

\bibitem[Weaire et~al.(2007)Weaire, Langlois, Saadatfar, and
  Hutzler]{Weaire:2007}
D.~Weaire, V.~Langlois, M.~Saadatfar, and S.~Hutzler.
\newblock \emph{Foam as granular matter}, volume~8 of \emph{Lecture Notes in
  Complex Systems}, chapter~1, pages 1--26.
\newblock World Scientific, 2007.
\newblock \doi{10.1142/9789812771995_0001}.

\bibitem[Weaire and Hutzler(2009)]{Weaire:2009}
D.~Weaire and S.~Hutzler.
\newblock Foam as a complex system.
\newblock \emph{J. Phys.: Condens. Matter}, 21:\penalty0 474227, 2009.
\newblock \doi{10.1088/0953-8984/21/47/474227}.

\bibitem[Odell et~al.(1981)Odell, Oster, Alberch, and Burnside]{Odell:1981}
G.M. Odell, G.~Oster, P.~Alberch, and B.~Burnside.
\newblock {The mechanical basis of morphogenesis: I. Epithelial folding and
  invagination}.
\newblock \emph{Dev. Biol.}, 85:\penalty0 446--462, 1981.
\newblock \doi{10.1016/0012-1606(81)90276-1}.

\bibitem[Kermode and Weaire(1990)]{Kermode:1990}
J.~P. Kermode and D.~Weaire.
\newblock {2D-FROTH: a program for the investigation of 2-dimensional froths}.
\newblock \emph{Comput. Phys. Commun.}, 60:\penalty0 75--109, 1990.
\newblock \doi{10.1016/0010-4655(90)90080-K}.

\bibitem[Ishimoto and Morishita(2014)]{Ishimoto:2014}
Y.~Ishimoto and Y.~Morishita.
\newblock {Bubbly vertex dynamics: A dynamical and geometrical model for
  epithelial tissues with curved cell shapes}.
\newblock \emph{Phys. Rev. E}, 90:\penalty0 052711, 2014.
\newblock \doi{10.1103/PhysRevE.90.052711}.

\bibitem[Perrone et~al.(2016)Perrone, Veldhuis, and Brodland]{Perrone:2016}
M.~C. Perrone, J.~H. Veldhuis, and G.~W. Brodland.
\newblock Non-straight cell edges are important to invasion and engulfment as
  demonstrated by cell mechanics model.
\newblock \emph{Biomech. Model. Mechanobiol.}, 15:\penalty0 405--418, 2016.
\newblock \doi{10.1007/s10237-015-0697-6}.

\bibitem[Boromand et~al.(2019)Boromand, Signoriello, Lowensohn, Orellana,
  Weeks, Ye, Shattuck, and O{'}Hern]{Boromand:2019}
A.~Boromand, A.~Signoriello, J.~Lowensohn, C.~S. Orellana, E.~R. Weeks, F.~Ye,
  M.~D. Shattuck, and C.~S. O{'}Hern.
\newblock The role of deformability in determining the structural and
  mechanical properties of bubbles and emulsions.
\newblock \emph{Soft Matter}, 15:\penalty0 5854--5865, 2019.
\newblock \doi{10.1039/C9SM00775J}.

\bibitem[Bolton and Weaire(1992)]{Bolton:1992}
F.~Bolton and D.~Weaire.
\newblock {The effects of Plateau borders in the two-dimensional soap froth.
  II. General simulation and analysis of rigidity loss transition}.
\newblock \emph{Phil. Mag. B}, 65:\penalty0 473--487, 1992.
\newblock \doi{10.1080/13642819208207644}.

\bibitem[Bolton and Dunne(1996)]{PLAT:1996}
F.~Bolton and F.~F. Dunne.
\newblock {Software PLAT: A computer code for simulating two-dimensional liquid
  foams}, 1996.
\newblock https://github.com/fbolton/plat.

\bibitem[Rejniak(2005)]{Rejniak:2005}
K.~A. Rejniak.
\newblock {A Single-Cell Approach in Modeling the Dynamics of Tumor
  Microregions}.
\newblock \emph{Math. Biosci. Eng.}, 2:\penalty0 643--655, 2005.
\newblock \doi{10.3934/mbe.2005.2.643}.

\bibitem[R. et~al.(2008)R., M., and K.]{Dillon:2008}
Dillon R., Owen M., and Painter K.
\newblock \emph{A single-cell based model of multicellular growth using the
  immersed boundary method}, pages 1--16.
\newblock Contemporary Mathematics. American Mathematical Society, 2008.
\newblock ISBN 9780821842676.

\bibitem[Jamali et~al.(2010)Jamali, Azimi, and Mofrad]{Jamali:2010}
Y.~Jamali, M.~Azimi, and M.~R.~K. Mofrad.
\newblock {A Sub-Cellular Viscoelastic Model for Cell Population Mechanics}.
\newblock \emph{PLOS ONE}, 5:\penalty0 1--20, 2010.
\newblock \doi{10.1371/journal.pone.0012097}.

\bibitem[{van der Sande} et~al.(2020){van der Sande}, Kraus, Houliston, and
  Kaandorp]{Sande:2020}
Maarten {van der Sande}, Yulia Kraus, Evelyn Houliston, and Jaap Kaandorp.
\newblock {A cell-based boundary model of gastrulation by unipolar ingression
  in the hydrozoan cnidarian Clytia hemisphaerica}.
\newblock \emph{Dev. Biol.}, 460:\penalty0 176--186, 2020.
\newblock \doi{10.1016/j.ydbio.2019.12.012}.

\bibitem[Tamulonis et~al.(2011)Tamulonis, Postma, Marlow, Magie, de~Jong, and
  Kaandorp]{Tamulonis:2011}
C.~Tamulonis, M.~Postma, H.~Q. Marlow, C.~R. Magie, J.~de~Jong, and
  J.~Kaandorp.
\newblock {A cell-based model of \textit{Nematostella vectensis} gastrulation
  including bottle cell formation, invagination and zippering}.
\newblock \emph{Dev. Biol.}, 351:\penalty0 217--228, 2011.
\newblock \doi{10.1016/j.ydbio.2010.10.017}.

\bibitem[Merks et~al.(2011)Merks, Guravage, Inz{\'e}, and Beemster]{Merks:2011}
R.~M.~H. Merks, M.~Guravage, D.~Inz{\'e}, and G.~T.~S. Beemster.
\newblock {VirtualLeaf: An Open-Source Framework for Cell-Based Modeling of
  Plant Tissue Growth and Development}.
\newblock \emph{Plant Physiol.}, 155:\penalty0 656--666, 2011.
\newblock \doi{10.1104/pp.110.167619}.

\bibitem[K\"ah\"ar\"a et~al.(2014)K\"ah\"ar\"a, Tallinen, and
  Timonen]{Kahara:2014}
T.~K\"ah\"ar\"a, T.~Tallinen, and J.~Timonen.
\newblock Numerical model for the shear rheology of two-dimensional wet foams
  with deformable bubbles.
\newblock \emph{Phys. Rev. E}, 90:\penalty0 032307, 2014.
\newblock \doi{10.1103/PhysRevE.90.032307}.

\bibitem[Mkrtchyan et~al.(2014)Mkrtchyan, \r{A}str\"{o}m, and
  Karttunen]{Mkrtchyan:2014}
A.~Mkrtchyan, J.~\r{A}str\"{o}m, and M.~Karttunen.
\newblock A new model for cell division and migration with spontaneous topology
  changes.
\newblock \emph{Soft Matter}, 10:\penalty0 4332--4339, 2014.
\newblock \doi{10.1039/C4SM00489B}.

\bibitem[Madhikar et~al.(2021)Madhikar, \AA{}str\"om, Baumeier, and
  Karttunen]{Madhikar:2021}
P.~Madhikar, J.~\AA{}str\"om, B.~Baumeier, and M.~Karttunen.
\newblock Jamming and force distribution in growing epithelial tissue.
\newblock \emph{Phys. Rev. Res.}, 3:\penalty0 023129, 2021.
\newblock \doi{10.1103/PhysRevResearch.3.023129}.

\bibitem[Tanaka et~al.(2015)Tanaka, Sichau, and Iber]{Tanaka:2015}
S.~Tanaka, D.~Sichau, and D.~Iber.
\newblock {LBIBCell: a cell-based simulation environment for morphogenetic
  problems}.
\newblock \emph{Bioinformatics}, 31:\penalty0 2340--2347, 2015.
\newblock \doi{10.1093/bioinformatics/btv147}.

\bibitem[Pitt-Francis et~al.(2009)Pitt-Francis, Pathmanathan, Bernabeu, Bordas,
  Cooper, Fletcher, Mirams, Murray, Osborne, Walter, Chapman, Garny, {van
  Leeuwen}, Maini, Rodr\'{i}guez, Waters, Whiteley, Byrne, and
  Gavaghan]{Pitt:2009}
J.~Pitt-Francis, P.~Pathmanathan, M.~O. Bernabeu, R.~Bordas, J.~Cooper, A.~G.
  Fletcher, G.~R. Mirams, P.~Murray, J.~M. Osborne, A.~Walter, S.~J. Chapman,
  A.~Garny, I.~M.~M. {van Leeuwen}, P.~K. Maini, B.~Rodr\'{i}guez, S.~L.
  Waters, J.~P. Whiteley, H.~M. Byrne, and D.~J. Gavaghan.
\newblock {Chaste: A test-driven approach to software development for
  biological modelling}.
\newblock \emph{Comput. Phys. Commun.}, 180:\penalty0 2452--2471, 2009.
\newblock \doi{10.1016/j.cpc.2009.07.019}.

\bibitem[Cooper et~al.(2017)Cooper, Baker, and Fletcher]{Cooper:2017}
F.~R. Cooper, R.~E. Baker, and A.~G. Fletcher.
\newblock {Numerical Analysis of the Immersed Boundary Method for Cell-Based
  Simulation}.
\newblock \emph{SIAM J. Sci. Comput.}, 39:\penalty0 B943--B967, 2017.
\newblock \doi{10.1137/16M1092246}.

\bibitem[Boromand et~al.(2018)Boromand, Signoriello, Ye, O'Hern, and
  Shattuck]{Boromand:2018}
A.~Boromand, A.~Signoriello, F.~Ye, C.~S. O'Hern, and M.~D. Shattuck.
\newblock {Jamming of Deformable Polygons}.
\newblock \emph{Phys. Rev. Lett.}, 121:\penalty0 248003, 2018.
\newblock \doi{10.1103/PhysRevLett.121.248003}.

\bibitem[Merchant(2016)]{numba-ncc:2016}
B.~Merchant.
\newblock {numba-ncc}, 2016.
\newblock https://github.com/bzm3r/numba-ncc.

\bibitem[Merchant(2020)]{rust-ncc:2020}
B.~Merchant.
\newblock {rust-ncc}, 2020.
\newblock https://github.com/bzm3r/rust-ncc.

\bibitem[Merchant et~al.(2018)Merchant, Edelstein-Keshet, and
  Feng]{Merchant:2018}
B.~Merchant, L.~Edelstein-Keshet, and J.~J. Feng.
\newblock {A Rho-GTPase based model explains spontaneous collective migration
  of neural crest cell clusters}.
\newblock \emph{Dev. Biol.}, 444:\penalty0 S262--S273, 2018.
\newblock \doi{10.1016/j.ydbio.2018.01.013}.

\bibitem[K\"{o}rner et~al.(2002)K\"{o}rner, Thies, and Singer]{Korner:2002}
C.~K\"{o}rner, M.~Thies, and R.~F. Singer.
\newblock {Modeling of Metal Foaming with Lattice Boltzmann Automata}.
\newblock \emph{Adv. Eng. Mater.}, 4:\penalty0 765--769, 2002.
\newblock \doi{10.1002/1527-2648(20021014)4:10<765::AID-ADEM765>3.0.CO;2-M}.

\bibitem[Latt et~al.(2020)Latt, Malaspinas, Kontaxakis, Parmigiani, Lagrava,
  Brogi, Belgacem, Thorimbert, Leclaire, Li, Marson, Lemus, Kotsalos, Conradin,
  Coreixas, Petkantchin, Raynaud, Beny, and Chopard]{Latt:2020}
J.~Latt, O.~Malaspinas, D.~Kontaxakis, A.~Parmigiani, D.~Lagrava, F.~Brogi,
  M.~Ben Belgacem, Y.~Thorimbert, S.~Leclaire, S.~Li, F.~Marson, J.~Lemus,
  C.~Kotsalos, R.~Conradin, C.~Coreixas, R.~Petkantchin, F.~Raynaud, J.~Beny,
  and B.~Chopard.
\newblock {Palabos: Parallel Lattice Boltzmann Solver}.
\newblock \emph{Comput. Math. Appl.}, 81:\penalty0 334--350, 2020.
\newblock \doi{10.1016/j.camwa.2020.03.022}.

\bibitem[Ataei et~al.(2021)Ataei, Shaayegan, Costa, Han, Park, and
  Bussmann]{Ataei:2021}
M.~Ataei, V.~Shaayegan, F.~Costa, S.~Han, C.~B. Park, and M.~Bussmann.
\newblock {LBfoam: An open-source software package for the simulation of
  foaming using the Lattice Boltzmann Method}.
\newblock \emph{Comput. Phys. Commun.}, 259:\penalty0 107698, 2021.
\newblock \doi{10.1016/j.cpc.2020.107698}.

\bibitem[Conradin et~al.(2021)Conradin, Coreixas, Latt, and
  Chopard]{Conradin:2021}
R.~Conradin, C.~Coreixas, J.~Latt, and B.~Chopard.
\newblock {PalaCell2D: A framework for detailed tissue morphogenesis}.
\newblock \emph{J. Comput. Sci.}, 53:\penalty0 101353, 2021.
\newblock \doi{10.1016/j.jocs.2021.101353}.

\bibitem[Brown et~al.(2021)Brown, Green, Binder, and Osborne]{Brown:2021}
P.~J. Brown, G.~E.~F. Green, B.~J. Binder, and J.~M. Osborne.
\newblock {A rigid body framework for multi-cellular modelling}.
\newblock \emph{Nat. Comput. Sci.}, 1:\penalty0 754--766, 2021.
\newblock \doi{10.1038/s43588-021-00154-4}.

\bibitem[Tervonen et~al.(2023)Tervonen, Korpela, Nymark, Hyttinen, and
  Ihalainen]{Tervonen:2023}
A.~Tervonen, S.~Korpela, S.~Nymark, J.~Hyttinen, and T.~O. Ihalainen.
\newblock {The Effect of Substrate Stiffness on Elastic Force Transmission in
  the Epithelial Monolayers over Short Timescales}.
\newblock \emph{Cell. Mol. Bioeng.}, 2023.
\newblock \doi{0.1007/s12195-023-00772-0}.

\bibitem[Nonomura(2012)]{Nonomura:2012}
M.~Nonomura.
\newblock {Study on Multicellular Systems Using a Phase Field Model}.
\newblock \emph{PLOS ONE}, 7:\penalty0 1--9, 2012.
\newblock \doi{10.1371/journal.pone.0033501}.

\bibitem[Palmieri et~al.(2015)Palmieri, Bresler, Wirtz, and
  Grant]{Palmieri:2015}
B.~Palmieri, Y.~Bresler, D.~Wirtz, and M.~Grant.
\newblock Multiple scale model for cell migration in monolayers: Elastic
  mismatch between cells enhances motility.
\newblock \emph{Sci. Rep.}, 5:\penalty0 11745, 2015.
\newblock \doi{10.1038/srep11745}.

\bibitem[L\"{o}ber et~al.(2015)L\"{o}ber, Ziebert, and Aranson]{Lober:2015}
J.~L\"{o}ber, F.~Ziebert, and I.~S. Aranson.
\newblock Collisions of deformable cells lead to collective migration.
\newblock \emph{Sci. Rep.}, 5:\penalty0 9172, 2015.
\newblock \doi{10.1038/srep09172}.

\bibitem[Biner(2017)]{Biner:2017}
S.~B. Biner.
\newblock \emph{{Programming Phase-Field Modeling}}.
\newblock Springer, 2017.
\newblock ISBN 978-3-319-41196-5.
\newblock \doi{10.1007/978-3-319-41196-5}.

\bibitem[Jiang et~al.(2019)Jiang, Garikipati, and Rudraraju]{Jiang:2019}
J.~Jiang, K.~Garikipati, and S~Rudraraju.
\newblock {A Diffuse Interface Framework for Modeling the Evolution of
  Multi-cell Aggregates as a Soft Packing Problem Driven by the Growth and
  Division of Cells}.
\newblock \emph{Bull. Math. Biol.}, 81:\penalty0 3282--3300, 2019.
\newblock \doi{10.1007/s11538-019-00577-1}.

\bibitem[Lavoratti et~al.(2021)Lavoratti, Heitkam, Hampel, and
  G.]{Lavoratti:2021}
T.~C. Lavoratti, S.~Heitkam, U.~Hampel, and Lecrivain G.
\newblock A computational method to simulate mono- and poly-disperse
  two-dimensional foams flowing in obstructed channel.
\newblock \emph{Rheol. Acta}, 60:\penalty0 587–--601, 2021.
\newblock \doi{10.1007/s00397-021-01288-y}.

\bibitem[Lecrivain(2021)]{Lecrivain:2021}
G.~Lecrivain.
\newblock {Data/Software for: Dynamics of mono- and poly-disperse
  two-dimensional foams flowing in an obstructed channel (Version 1.0)}, 2021.
\newblock Rodare.

\bibitem[Jantsch-Plunger and Glotzer(1999)]{Jantsch:1999}
V.~Jantsch-Plunger and M.~Glotzer.
\newblock Depletion of syntaxins in the early \textit{Caenorhabditis elegans}
  embryo reveals a role for membrane fusion events in cytokinesis.
\newblock \emph{Curr. Biol.}, 9:\penalty0 738--745, 1999.
\newblock \doi{10.1016/S0960-9822(99)80333-9}.

\bibitem[Lu and Kang(2009)]{Lu:2009}
X.~Lu and Y.~Kang.
\newblock {Cell Fusion as a Hidden Force in Tumor Progression}.
\newblock \emph{Cancer Research}, 69:\penalty0 8536--8539, 2009.
\newblock \doi{10.1158/0008-5472.CAN-09-2159}.

\bibitem[Rochlin et~al.(2010)Rochlin, Yu, Roy, and Baylies]{Rochlin:2010}
K.~Rochlin, S.~Yu, S.~Roy, and M.~K. Baylies.
\newblock {Myoblast fusion: When it takes more to make one}.
\newblock \emph{Dev. Biol.}, 341:\penalty0 66--83, 2010.
\newblock \doi{10.1016/j.ydbio.2009.10.024}.

\bibitem[Stillwell(2016)]{Stillwell:2016}
W.~Stillwell.
\newblock \emph{Moving Components Through the Cell: Membrane Trafficking},
  chapter~17, pages 369--379.
\newblock Elsevier, 2 edition, 2016.
\newblock ISBN 978-0-444-63772-7.
\newblock \doi{10.1016/B978-0-444-63772-7.00017-8}.

\bibitem[Bergou et~al.(2008)Bergou, Wardetzky, Robinson, Audoly, and
  Grinspun]{Bergou:2008}
M.~Bergou, M.~Wardetzky, S.~Robinson, B.~Audoly, and E.~Grinspun.
\newblock Discrete elastic rods.
\newblock In \emph{ACM SIGGRAPH 2008 Papers}, SIGGRAPH '08, page~63, 2008.
\newblock \doi{10.1145/1399504.1360662}.

\bibitem[Soerjadi(1968)]{Soerjadi:1968}
R.~Soerjadi.
\newblock {On the Computation of the Moments of a Polygon, with some
  Applications}.
\newblock \emph{Heron}, 16:\penalty0 43--58, 1968.

\bibitem[Vetter et~al.(2013)Vetter, Wittel, Stoop, and Herrmann]{Vetter:2013}
R.~Vetter, F.~K. Wittel, N.~Stoop, and H.~J. Herrmann.
\newblock Finite element simulation of dense wire packings.
\newblock \emph{Eur. J. Mech. A Solids}, 37:\penalty0 160--171, 2013.
\newblock \doi{10.1016/j.euromechsol.2012.06.007}.

\bibitem[Knill(2004)]{Knill:2004}
O.~Knill.
\newblock Eigenvalues and eigenvectors of 2x2 matrices.
\newblock
  \url{https://people.math.harvard.edu/~knill/teaching/math21b2004/exhibits/2dmatrices/index.html},
  2004.
\newblock Harvard University.

\bibitem[Franklin(2006)]{Franklin:2006}
W.~R. Franklin.
\newblock {PNPOLY - Point Inclusion in Polygon Test}.
\newblock \url{https://wrfranklin.org/Research/Short_Notes/pnpoly.html}, 2006.

\bibitem[Quentrec and Brot(1973)]{Quentrec:1973}
B.~Quentrec and C.~Brot.
\newblock New method for searching for neighbors in molecular dynamics
  computations.
\newblock \emph{J. Comput. Phys.}, 13:\penalty0 430--432, 1973.
\newblock \doi{10.1016/0021-9991(73)90046-6}.

\bibitem[Ziebert and Aranson(2016)]{Ziebert:2016}
F.~Ziebert and I.~Aranson.
\newblock Computational approaches to substrate-based cell motility.
\newblock \emph{npj Comput. Mater.}, 2:\penalty0 16019, 2016.
\newblock \doi{10.1038/npjcompumats.2016.19}.

\bibitem[Treado et~al.(2021)Treado, Wang, Boromand, Murrell, Shattuck, and
  O'Hern]{Treado:2021}
J.~D. Treado, D.~Wang, A.~Boromand, M.~P. Murrell, M.~D. Shattuck, and C.~S.
  O'Hern.
\newblock Bridging particle deformability and collective response in soft
  solids.
\newblock \emph{Phys. Rev. Mater.}, 5:\penalty0 055605, 2021.
\newblock \doi{10.1103/PhysRevMaterials.5.055605}.

\bibitem[Sapala et~al.(2018)Sapala, Runions, Routier-Kierzkowska, Das~Gupta,
  Hong, Hofhuis, Verger, Mosca, Li, Hay, Hamant, Roeder, Tsiantis,
  Prusinkiewicz, and Smith]{Sapala:2018}
A.~Sapala, A.~Runions, A.-L. Routier-Kierzkowska, M.~Das~Gupta, L.~Hong,
  H.~Hofhuis, S.~Verger, G.~Mosca, C.-B. Li, A.~Hay, O.~Hamant, A.~H.~K.
  Roeder, M.~Tsiantis, P.~Prusinkiewicz, and R.~S. Smith.
\newblock Why plants make puzzle cells, and how their shape emerges.
\newblock \emph{eLife}, 7:\penalty0 e32794, 2018.
\newblock \doi{10.7554/eLife.32794}.

\bibitem[Stoop et~al.(2008)Stoop, Wittel, and Herrmann]{Stoop:2008}
N.~Stoop, F.~K. Wittel, and H.~J. Herrmann.
\newblock {Morphological Phases of Crumpled Wire}.
\newblock \emph{Phys. Rev. Lett.}, 101:\penalty0 094101, 2008.
\newblock \doi{10.1103/PhysRevLett.101.094101}.

\bibitem[Halver and Sutmann(2016)]{Halver:2016}
R.~Halver and G.~Sutmann.
\newblock {Multi-threaded Construction of Neighbour Lists for Particle Systems
  in OpenMP}.
\newblock In R.~Wyrzykowski, E.~Deelman, J.~Dongarra, K.~Karczewski,
  J.~Kitowski, and K.~Wiatr, editors, \emph{Parallel Processing and Applied
  Mathematics}, volume 9574 of \emph{Lecture Notes in Computer Science}, pages
  153--165, 2016.
\newblock \doi{10.1007/978-3-319-32152-3_15}.

\bibitem[Runser et~al.(2023)Runser, Vetter, and Iber]{Runser:2023}
S.~Runser, R.~Vetter, and D.~Iber.
\newblock {3D Simulation of Tissue Mechanics with Cell Polarization}.
\newblock \emph{BioRxiv}, 2023.
\newblock \doi{10.1101/2023.03.28.534574}.

\bibitem[M.~Phillips(2007)]{OFF}
T.~Munzner M.~Phillips, S.~Levy.
\newblock \emph{GeomView Manual}.
\newblock The Geometry Center, University of Minnesota, 2007.
\newblock URL \url{http://www.geomview.org/docs/geomview.pdf}.
\newblock Section 4.2.5.

\bibitem[Inc.(2010)]{VTK}
Kitware Inc.
\newblock \emph{The VTK User’s Guide}.
\newblock Kitware, 11th edition, 2010.
\newblock ISBN 978-1-930934-23-8.
\newblock URL \url{https://www.kitware.com/products/books/VTKUsersGuide.pdf}.
\newblock Section 19.3.

\bibitem[G\'{o}mez et~al.(2021)G\'{o}mez, Dumond, Hodel, Vetter, and
  Iber]{Gomez:2021}
H.~F. G\'{o}mez, Mathilde~S. Dumond, L.~Hodel, R.~Vetter, and D.~Iber.
\newblock 3d cell neighbour dynamics in growing pseudostratified epithelia.
\newblock \emph{eLife}, 10:\penalty0 e68135, 2021.
\newblock \doi{10.7554/eLife.68135}.

\bibitem[Iber and Vetter(2022{\natexlab{a}})]{Iber:2022a}
D.~Iber and R.~Vetter.
\newblock 3d organisation of cells in pseudostratified epithelia.
\newblock \emph{Front. Phys.}, 10:\penalty0 898160, 2022{\natexlab{a}}.
\newblock \doi{10.3389/fphy.2022.898160}.

\bibitem[Iber and Vetter(2022{\natexlab{b}})]{Iber:2022b}
D.~Iber and R.~Vetter.
\newblock Relationship between epithelial organization and morphogen
  interpretation.
\newblock \emph{Curr. Opin. Genet. Dev.}, 75:\penalty0 101916,
  2022{\natexlab{b}}.
\newblock \doi{10.1016/j.gde.2022.101916}.

\bibitem[Spear and Erickson(2012)]{Spear:2012}
Philip~C. Spear and Carol~A. Erickson.
\newblock {Interkinetic nuclear migration: A mysterious process in search of a
  function}.
\newblock \emph{Develop. Growth Differ.}, 54:\penalty0 306--316, 2012.
\newblock \doi{10.1111/j.1440-169X.2012.01342.x}.

\bibitem[Buchmann et~al.(2014)Buchmann, Alber, and Zartman]{Buchmann:2014}
A.~Buchmann, M.~Alber, and J.~J. Zartman.
\newblock {Sizing it up: The mechanical feedback hypothesis of organ growth
  regulation}.
\newblock \emph{Semin. Cell Dev. Biol.}, 35:\penalty0 73--81, 2014.
\newblock \doi{10.1016/j.semcdb.2014.06.018}.

\bibitem[Wu et~al.(2020)Wu, Gilkes, Phillip, Narkar, Cheng, Marchand, Lee, Li,
  and Wirtz]{Wu:2020}
P.-H. Wu, D.~M. Gilkes, J.~M. Phillip, A.~Narkar, T.~W.-T. Cheng, J.~Marchand,
  M.-H. Lee, R.~Li, and D.~Wirtz.
\newblock Single-cell morphology encodes metastatic potential.
\newblock \emph{Sci. Adv.}, 6:\penalty0 eaaw6938, 2020.
\newblock \doi{10.1126/sciadv.aaw6938}.

\bibitem[Hern\'{a}ndez and Podbilewicz(2017)]{Hernandez:2017}
J.~M. Hern\'{a}ndez and B.~Podbilewicz.
\newblock {The hallmarks of cell-cell fusion}.
\newblock \emph{Development}, 144:\penalty0 4481--4495, 2017.
\newblock \doi{10.1242/dev.155523}.

\bibitem[Krajnc et~al.(2021)Krajnc, Stern, and Zankoc]{Krajnc:2021}
M.~Krajnc, T.~Stern, and C.~Zankoc.
\newblock Active instability and nonlinear dynamics of cell-cell junctions.
\newblock \emph{Phys. Rev. Lett.}, 127:\penalty0 198103, 2021.
\newblock \doi{10.1103/PhysRevLett.127.198103}.

\bibitem[Verlet(1967)]{Verlet:1967}
L.~Verlet.
\newblock {Computer ``Experiments'' on Classical Fluids. I. Thermodynamical
  Properties of Lennard-Jones Molecules}.
\newblock \emph{Phys. Rev.}, 159:\penalty0 98--103, 1967.
\newblock \doi{10.1103/PhysRev.159.98}.

\bibitem[Da et~al.(2014)Da, Barry, and Grinspun]{Da:2014}
F.~Da, C.~Barry, and E.~Grinspun.
\newblock {Multimaterial mesh-based surface tracking}.
\newblock \emph{ACM Trans. Graph.}, 33:\penalty0 1--11, 2014.
\newblock \doi{10.1145/2601097.2601146}.

\bibitem[Madhikar et~al.(2018)Madhikar, \r{A}str\"{o}m, Westerholm, and
  Karttunen]{Madhikar:2018}
P.~Madhikar, J.~\r{A}str\"{o}m, J.~Westerholm, and M.~Karttunen.
\newblock {CellSim3D: GPU accelerated software for simulations of cellular
  growth and division in three dimensions}.
\newblock \emph{Comput. Phys. Commun.}, 232:\penalty0 206--213, 2018.
\newblock \doi{10.1016/j.cpc.2018.05.024}.

\bibitem[Van~Liedekerke et~al.(2020)Van~Liedekerke, Neitsch, Johann, Warmt,
  Gonz\`{a}lez-Valverde, Hoehme, Grosser, Kaes, and Drasdo]{Liedekerke:2020}
P.~Van~Liedekerke, J.~Neitsch, T.~Johann, E.~Warmt, I.~Gonz\`{a}lez-Valverde,
  S.~Hoehme, S.~Grosser, J.~Kaes, and D.~Drasdo.
\newblock A quantitative high-resolution computational mechanics cell model for
  growing and regenerating tissues.
\newblock \emph{Biomech. Model. Mechanobiol.}, 19:\penalty0 189--220, 2020.
\newblock \doi{10.1007/s10237-019-01204-7}.

\bibitem[Wang et~al.(2021)Wang, Treado, Boromand, Norwick, Murrell, Shattuck,
  and O'Hern]{Wang:2021}
D.~Wang, J.~D. Treado, A.~Boromand, B.~Norwick, M.~P. Murrell, M.~D. Shattuck,
  and C.~S. O'Hern.
\newblock The structural, vibrational, and mechanical properties of jammed
  packings of deformable particles in three dimensions.
\newblock \emph{Soft Matter}, 17:\penalty0 9901--9915, 2021.
\newblock \doi{10.1039/D1SM01228B}.

\bibitem[Torres-S\'{a}nchez et~al.(2022)Torres-S\'{a}nchez, Kerr~Winter, and
  Salbreux]{Torres:2022}
A.~Torres-S\'{a}nchez, M.~Kerr~Winter, and G.~Salbreux.
\newblock {Interacting active surfaces: A model for three-dimensional cell
  aggregates}.
\newblock \emph{PLoS Comput. Biol.}, 18:\penalty0 e1010762, 2022.
\newblock \doi{10.1371/journal.pcbi.1010762}.

\bibitem[Okuda and Hiraiwa(2023)]{Okuda:2023}
S.~Okuda and T.~Hiraiwa.
\newblock Modelling contractile ring formation and division to daughter cells
  for simulating proliferative multicellular dynamics.
\newblock \emph{Eur. Phys. J. E}, 46:\penalty0 56, 2023.
\newblock \doi{10.1140/epje/s10189-023-00315-5}.

\end{thebibliography}
\end{document}